\documentclass[10pt]{emulateapj}
\usepackage{amssymb}
\usepackage{amsbsy}
\usepackage{amsfonts}
\usepackage{pslatex}
\usepackage{graphicx}

\begin{document}

\title{Perpendicular Ion Heating by Low-Frequency Alfv\'en-Wave Turbulence in the Solar Wind}

\author{Benjamin D. G. Chandran\altaffilmark{1}, Bo Li\altaffilmark{2}, Barrett N. Rogers\altaffilmark{2}, Eliot Quataert\altaffilmark{3}, \& Kai Germaschewski\altaffilmark{1}}

\altaffiltext{1}{Space Science Center and Department of Physics, University of New Hampshire, Durham, NH; benjamin.chandran@unh.edu, kai.germaschewski@unh.edu}

\altaffiltext{2}{Department of Physics \& Astronomy, Dartmouth College, Hanover, NH; 
bo.li.physics@dartmouth.edu, rogers@endurance.dartmouth.edu}

\altaffiltext{3}{Astronomy Department \& Theoretical Astrophysics Center, 601 Campbell Hall, The University of California, Berkeley, CA 94720; eliot@astro.berkeley.edu} 

\begin{abstract}
  We consider ion heating by turbulent Alfv\'en waves (AWs) and
  kinetic Alfv\'en waves (KAWs) with wavelengths (measured
  perpendicular to the magnetic field) that are comparable to the ion
  gyroradius and frequencies~$\omega$ smaller than the ion cyclotron
  frequency~$\Omega $.  As in previous studies, we find that when the
  turbulence amplitude exceeds a certain threshold, an ion's orbit
  becomes chaotic. The ion then interacts stochastically with the
  time-varying electrostatic potential, and the ion's energy undergoes
  a random walk. Using phenomenological arguments, we derive an
  analytic expression for the rates at which different ion species are
  heated, which we test by simulating test particles interacting with
  a spectrum of randomly phased AWs and KAWs. We find that the
  stochastic heating rate depends sensitively on the quantity
  $\epsilon = \delta v_{\rho}/v_\perp $, where $v_\perp$
  ($v_\parallel$) is the component of the ion velocity perpendicular
  (parallel) to the background magnetic field~${\bf B}_0$, and $\delta
  v_\rho$ ($\delta B_\rho$) is the rms amplitude of the velocity
  (magnetic-field) fluctuations at the gyroradius scale. In the case
  of thermal protons, when $\epsilon \ll \epsilon_{\rm crit}$, where
  $\epsilon _{\rm crit}$ is a dimensionless constant, a proton's
  magnetic moment is nearly conserved and stochastic heating is
  extremely weak.  However, when $\epsilon > \epsilon_{\rm
    crit}$, the proton heating rate exceeds the cascade power that
  would be present in strong balanced KAW turbulence with the same
  value of~$\delta v_\rho$, and magnetic-moment conservation is
  violated even when $\omega \ll \Omega$.  For the random-phase waves
  in our test-particle simulations, $\epsilon_{\rm crit} \simeq 0.2$.
  For protons in low-$\beta$ plasmas, $\epsilon \simeq \beta^{-1/2}
  \delta B_\rho/B_0$, and~$\epsilon$ can exceed~$\epsilon_{\rm crit}$
  even when $\delta B_\rho/B_0\ll \epsilon_{\rm crit}$, where~$\beta$
  is the ratio of plasma pressure to magnetic pressure.  The heating
  is anisotropic, increasing~$v_\perp^2$ much more
  than~$v_\parallel^2$ when $\beta \ll 1$.  (In contrast, at $\beta
  \gtrsim 1$ Landau damping and transit-time damping of KAWs lead to
  strong parallel heating of protons.)  At comparable temperatures,
  alpha particles and minor ions have larger values of~$\epsilon$ than
  protons and are heated more efficiently as a result.  We discuss the
  implications of our results for ion heating in coronal holes and the
  solar wind.
\end{abstract}


\maketitle

\section{Introduction}
\label{intro}

Beginning in the 1960s, a number of authors developed steady-state
hydrodynamic models of the solar wind, in which the temperature was fixed
at the coronal base and the solar wind was heated by thermal
conduction~(e.g. Parker 1965; Hartle \& Sturrock 1968; Durney 1972; Holzer
\& Leer~1980). \nocite{parker65,hartle68,durney72,holzer80} For realistic
values of the coronal temperature and density, these models were unable to
reproduce the large flow velocities of fast-solar-wind streams at 1~AU,
suggesting that the fast wind is heated above the coronal base by
some additional mechanism. Observational evidence for extended,
non-conductive heating has since been provided by measurements from the
Ultraviolet Coronagraph Spectrometer (UVCS), which show radially increasing
minor-ion temperatures in coronal holes (the open-magnetic-field-line regions from which the fast wind emanates) at heliocentric distances~$r$
between $1.5 R_{\sun}$ and $3.5 R_{\sun}$ \citep{kohl98,antonucci00}.
Identifying the physical mechanisms responsible for this heating and
determining the heating rates of the different particle species are among
the major challenges in the study of the solar wind at the present time.

One of the first mechanisms proposed to account for solar-wind heating was turbulence~\citep{coleman68}. The
importance of turbulent heating is suggested by {\em in situ} measurements
of ubiquitous, large-amplitude fluctuations in the velocity and magnetic
field in the interplanetary medium~\citep{belcher71,goldstein95,bruno05},
as well as the positive correlation between the solar-wind temperature and
the amplitude of the fluctuations~\citep{grappin90,vasquez07}. In addition,
 the expected rate at which the measured
fluctuations dissipate (based on phenomenological turbulence theories) is
comparable to the observationally inferred solar-wind heating
rate~\citep{smith01,breech09,cranmer09}. The {\em in situ} measurements on
which the above studies are based are limited to the locations where
spacecraft have flown - that is, to $r\gtrsim 0.3$~AU.  However, the
velocity and magnetic-field fluctuations are often correlated in the sense
of Alfv\'en waves propagating away from the Sun in the solar-wind
frame~\citep{belcher71,tumarsch95,bavassano00}, indicating that these waves
originate at or near the Sun, consistent with the idea that turbulent
heating remains important as $r$ decreases below~$0.3$~AU.

At least two different scenarios for turbulent heating of coronal
holes and the solar wind are possible. In the first, magnetic
reconnection or some other process launches Alfv\'en waves into the
corona, including waves with $|k_\parallel | \gtrsim k_\perp$, where
$k_\parallel$ and $k_\perp$ are the components of the wavevector~${\bf
  k}$ parallel and perpendicular to the local background magnetic
field~${\bf B}_0$.\footnote{Alfv\'en waves play a key role in
  extended-heating models because fast magnetosonic waves can not in
  general escape from the chromosphere into the corona since they are
  reflected at the transition region~\citep{hollweg78b}. In addition,
  slow magnetosonic waves are strongly damped in collisionless
  low-$\beta$ plasmas \citep{barnes66} and thus are not an effective vehicle for
  transporting energy from the coronal base to~$r\gtrsim 1.5
  R_{\sun}$.}  Given the large Alfv\'en speed~$v_{\rm A}$ in coronal
holes ($\gtrsim 10^3$~km/s), the frequency~$\omega = k_\parallel
v_{\rm A}$ of such waves exceeds 1~Hz for wavelengths shorter
than~$\sim 10^4$~km. Once such waves enter the corona, nonlinear
interactions with coronal density fluctuations [which are inferred
from radio observations~\citep{coles89}] can convert a significant
fraction of the Alfv\'en wave power into fast magnetosonic
waves~\citep{chandran08b}. The energy in fast magnetosonic waves can
then cascade to higher frequencies~\citep{cho02,svidzinski10}, generating
high-frequency Alfv\'en waves with $k_\parallel > k_\perp$ in the
process~\citep{chandran05a}.  Although the dissipation of
high-frequency fast waves and Alfv\'en/ion-cyclotron waves could in
principle explain the UVCS observations of ion
heating~\citep{lih01,hollweg02,markovskii10a}, there is no direct
observational evidence that waves with high frequencies and/or
$|k_\parallel| \gtrsim k_\perp$ are present in coronal holes.

An alternative scenario, which we focus on in this paper, involves the
launching of much lower-frequency Alfv\'en waves by convective photospheric
motions. An effective $k_\perp$ for such waves can be estimated as
$2\pi/L_0$, where $L_0$ is of order the average spacing of either supergranules ($\sim
3\times 10^4$~km) or photospheric flux tubes ($\sim 5000$~km). For a wave
period~$2\pi/(k_\parallel v_{\rm A})$ of $10^3$~s and an Alfv\'en speed
of~$10^3$~km/s, the ratio $|k_\parallel|/k_\perp$ of such waves is~$\leq
0.03$.  Such highly anisotropic Alfv\'en waves are inefficient at
generating compressive modes~\citep{cho03a,chandran05a,chandran08b}. On the
other hand, they can interact with oppositely propagating Alfv\'en waves,
causing wave energy to cascade from large scales to small scales, or,
equivalently, small~$k$ to large~$k$, where the fluctuations dissipate,
heating the ambient plasma~\citep{iroshnikov63,kraichnan65}. Although the
Sun launches only outward-propagating Alfv\'en waves, the
inward-propagating waves required for the Alfv\'en-wave cascade are
generated near the Sun by non-WKB wave reflection arising from the gradient
in the Alfv\'en speed~\citep{heinemann80,velli89,matthaeus99b,dmitruk02,cranmer05,verdini07,hollweg07,verdini09b}.
An important development in the theory of Alfv\'en-wave turbulence was the
discovery that interactions between oppositely propagating Alfv\'en waves
cause wave energy to cascade primarily to larger~$k_\perp$ and only weakly
to larger~$|k_\parallel|$~\citep{montgomery81,shebalin83,goldreich95}. At
the dissipation scale, the value of $|k_\parallel|/k_\perp$ is thus even
smaller than at the driving scale~$L_0$.

This second scenario for solar-wind heating is compelling for a number of reasons. For example, convective photospheric motions inevitably launch low-frequency Alfv\'en waves into the corona by perturbing the footpoints of open magnetic field lines, and low-frequency Alfv\'en waves are observed in the corona~\citep{tomczyk07} and at $r>0.3$~AU~\citep{belcher71}. In addition, several models have been developed to describe wave reflection and turbulent heating by low-frequency Alfv\'en waves in the fast solar wind, taking into account the solar-wind velocity, density, and magnetic-field profiles, and incorporating observational constraints on the Alfv\'en-wave amplitudes; in all of these models, the turbulent heating rate appears to be consistent with the requirements for generating the fast wind~\citep{cranmer05,cranmer07,chandran09c,verdini09a,verdini10}. We also note that radio observations of density fluctuations provide an upper limit on the heating rate from Alfv\'en waves in coronal holes, since the Alfv\'en waves become increasingly compressive with increasing~$k$. Although these upper limits rule out fast-wind generation by non-turbulent high-frequency Alfv\'en/ion-cyclotron waves (unless ${\bf k}$ is nearly parallel to~${\bf B}_0$ for all the waves; Hollweg~2000), \nocite{hollweg00} they are consistent with fast-wind generation by low-frequency (kinetic) Alfv\'en-wave turbulence with $k_\perp \gg k_\parallel$~\citep{harmon05,chandran09d}.

Despite these considerations, it is not clear that low-frequency
Alfv\'en-wave turbulence can explain two key observations. First,
measurements of the proton and electron temperature profiles in the
fast solar wind at $r>0.3$~AU demonstrate that the proton heating rate
exceeds the electron heating rate by a modest
factor~\citep{cranmer09}. Similarly, empirically constrained fluid
models of coronal holes including thermal conduction suggest that
protons receive a substantial fraction ($\sim 0.5$) of the total
heating power~\citep{allen98}. Second, UVCS observations show that
minor ions such as~${\rm O}^{+5}$ are heated in such a way that
thermal motions perpendicular to ${\bf B}_0$ are much more rapid than
thermal motions along~${\bf B}_0$ (i.e., $T_\perp \gg
T_\parallel$)~\citep{kohl98,antonucci00}.  A similar temperature
anisotropy is measured {\em in situ} at $r>0.3$~AU for protons in
fast-solar-wind streams with $\beta \ll 1$, despite the fact that
(double) adiabatic expansion acts to
decrease~$T_\perp/T_\parallel$~\citep{marsch82,marsch04,hellinger06},
where $\beta = 8\pi p/B^2$ is the ratio of the plasma pressure to the
magnetic pressure.  Thus, in coronal holes and fast wind with $\beta
\ll 1$, ions receive $\gtrsim 1/2$ of the total heating, and ion
heating is mostly ``perpendicular to the magnetic field.''

Because the rms amplitude of the magnetic-field fluctuation $\delta B$
at the dissipation scale is $\ll B_0$, the damping of turbulent
fluctuations can be treated, to a first approximation, using the
Vlasov-Maxwell theory of linear waves. In this theory, Alfv\'en waves
are virtually undamped when $k_\perp \rho_{\rm p} \ll 1$ and $\omega
\ll \Omega_{\rm p}$, where $\rho_{\rm p}$ is the rms proton gyroradius
and $\Omega_{\rm p}$ is the proton cyclotron frequency. However, as
$k_\perp \rho_{\rm p}$ increases to values $\gtrsim 1$, the Alfv\'en
waves (AWs) become kinetic Alfv\'en waves (KAWs), the ions begin to
decouple from the electrons, and the waves develop fluctuating
electric-field and magnetic-field components parallel to~${\bf
  B}_0$~\citep{hasegawa76a,schekochihin09}.  For KAWs with
$k_\parallel \ll k_\perp$ and $ \omega \ll \Omega_{\rm p}$, the
primary damping mechanisms are Landau damping and/or transit-time
damping, which lead to parallel heating of the plasma, not
perpendicular heating~\citep{quataert98,leamon99,cranmer03,gary04a}.
Moreover, in low-$\beta$ plasmas, the waves damp almost entirely on
the electrons, because thermal ions are too slow to satisfy the Landau
resonance condition~$\omega - k_\parallel v_\parallel =
0$~\citep{quataert98,gruzinov98}.  Thus, if KAWs damp according to
linear Vlasov theory, then they are unable to explain the strong
perpendicular ion heating that is inferred from observations.  This
discrepancy casts doubt on the viability of low-frequency AW/KAW
turbulence as a mechanism for heating coronal holes and the fast solar
wind.

A number of studies have gone beyond linear Vlasov theory to
investigate the possibility of perpendicular ion heating by
low-frequency AW/KAW turbulence.  \cite{johnson01}, \cite{chen01},
\cite{white02}, and \cite{voitenko04} investigated the dissipation of
mono-chromatic KAWs and AWs with $\omega < \Omega_{\rm p}$, finding
that such waves cause perpendicular ion heating if the wave amplitude
exceeds a minimum threshold.  \cite{dmitruk04} and \cite{lehe09}
simulated test particles propagating in the electric and magnetic
fields resulting from direct numerical simulations of
magnetohydrodynamic (MHD) turbulence at $0.1 \lesssim \beta \lesssim
10$.  They both found perpendicular ion heating under some conditions,
but \cite{lehe09} argued that the perpendicular heating seen in both
studies is due to cyclotron resonance and does not apply to the solar
wind because it is an artifact of limited numerical resolution.
\cite{parashar09} found perpendicular ion heating in two-dimensional
hybrid simulations of a turbulent plasma, in which ions are treated as
particles and electrons are treated as a fluid. In addition,
\cite{markovskii02b} and \cite{markovskii06} investigated
high-frequency secondary instabilities that are generated by KAWs near
the gyroradius scale, and argued that such instabilities may be able
to explain the observed perpendicular ion heating.

In this paper, we continue this general line of inquiry and address an
important open problem: determining the perpendicular ion heating rate
in anisotropic, low-frequency ($\omega < \Omega_{\rm p}$), AW/KAW
turbulence as a function of the amplitude of the turbulent
fluctuations at the gyroradius scale.  In section~\ref{sec:stoch} we
develop a phenomenological theory of stochastic ion heating, obtaining
an approximate analytic expression for the heating rates of different
ion species. We also present simulations of test particles propagating
in a spectrum of AWs and KAWs to test our phenomenological theory and
to determine the two dimensionless constants that
appear in our expression for the heating rate. In section~\ref{sec:sw}
we apply our results to perpendicular ion heating in coronal holes and
the fast solar wind.

\section{Stochastic Ion Heating by Alfv\'enic Turbulence at the Gyroradius Scale}
\label{sec:stoch} 

We consider ion heating by fluctuations with
transverse length scales~$\lambda_\perp$ (measured perpendicular
to~${\bf B}_0$) of order the ion gyroradius~$\rho  = v_\perp
/\Omega  $ (i.e., $k_\perp \rho  \sim 1$), where
$\Omega  = qB_0/m c$ is the ion cyclotron frequency, and $q$ and
$m$ are the ion charge and mass. We assume that $\rho  \gtrsim
\rho_{\rm p}$, where 
\begin{equation}
\rho_{\rm p } = \frac{v_{\perp {\rm p}}}{\Omega_{\rm p}}
\label{eq:orhop} 
\end{equation} 
is the rms proton gyroradius in the background magnetic field,
\begin{equation}
v_{\perp {\rm p}} = \sqrt{\frac{2k_{\rm B} T_{\rm p}}{m_{\rm p}}}
\label{eq:vperpp} 
\end{equation} 
is the rms perpendicular velocity of protons,  $T_{\rm p}$ is the
(perpendicular) proton temperature, and $m_{\rm p}$ is the proton mass.
If $\rho  \gg \rho _{\rm p}$, then the gyro-scale
fluctuations are AWs. If $\rho  \sim \rho_{\rm p}$,
then the gyro-scale fluctuations are KAWs.

We define $\delta v_\rho$ and $\delta B_\rho$ to be the rms
amplitudes of the fluctuating velocity and magnetic-field vectors at
$k_\perp \rho  \sim 1$.  Similarly, $\delta E_\rho $ and $\delta
\Phi_\rho$ are the rms amplitudes of the fluctuating electric field and
electrostatic potential at $k_\perp \rho  \sim 1$.  We assume that
$\delta v_\rho$, $\delta B_\rho$, $\delta E_\rho$, and $\delta \Phi_\rho$
are related to one another in the same way that the magnitudes of the
fluctuating velocity, magnetic field, electric field, and electrostatic
potential are related in a linear (kinetic) Alfv\'en wave.  Thus, since
$k_\perp \rho_{\rm p} \lesssim 1$,
\begin{equation}
\delta E_\rho \simeq \frac{\delta v_\rho B_0}{c},
\label{eq:E1} 
\end{equation} 
$\delta \Phi_\rho \sim \rho  \delta E_\rho$, and 
\begin{equation}
q \,\delta \Phi_\rho \sim m v_\perp \,\delta v_\rho.
\label{eq:dPhi1} 
\end{equation} 
The fractional change in an ion's perpendicular kinetic energy~$m v_\perp^2/2$ during
a single gyro-period is then given by
\begin{equation}
\frac{2 q\, \delta \Phi_\rho}{m v_\perp^2} \sim 2 \epsilon,
\end{equation} 
where
\begin{equation}
\epsilon = \frac{\delta v_\rho}{v_\perp}.
\label{eq:defeps} 
\end{equation} 

When $\epsilon \ll 1$, an ion's kinetic energy is nearly constant
during a single gyro-period. If in addition $\delta B_\rho \ll B_0$,
then the ion's orbit in the plane perpendicular to~${\bf B}_0$ closely
approximates a closed circle in some suitably chosen reference
frame. In this case, the ion possesses an adiabatic invariant of the
form $J = \oint p dq$ that is conserved to a high degree of accuracy,
where $q$ is the angular coordinate corresponding to the particle's
nearly periodic cyclotron gyration and $p$ is the canonically
conjugate momentum~\citep{kruskal62}. In the limit of
small~$\epsilon$, $J$ is approximately equal to the magnetic
moment~$\mu = mv_\perp^2/2B$. The near conservation of~$J$ implies
that perpendicular ion heating is extremely weak.  In
Appendix~\ref{ap:mu}, we present a calculation for electrostatic waves
with $k_\perp \rho  \sim 1$ and $\epsilon \ll 1$ that
illustrates how the leading-order terms in the time derivative of
$v_\perp^2$ are unable to cause secular growth in~$T_\perp$.

On the other hand, as $\epsilon$ increases from 0 to~1, the fractional
change in an ion's perpendicular kinetic energy during a single
gyro-period grows to a value of order unity. We treat the spatial
variations in the electrostatic potential~$\Phi$ at $k_\perp \rho \sim
1$ as random or disordered, as is the case in turbulence or a spectrum
of many randomly phased waves.  Thus, when $\epsilon$ exceeds some
threshold (whose value we investigate below), an ion's orbit in the
plane perpendicular to~${\bf B}_0$ becomes chaotic. In this case, the
ion's orbit does not satisfy the criteria for the approximate
conservation of~$J$~\citep{kruskal62}, and perpendicular ion heating
becomes possible~\citep{johnson01,chen01,white02}.

To estimate the rate at which ions are heated, we begin by considering the
 Hamiltonian of a particle of charge~$q$ and mass~$m$,
\begin{equation}
H = q\Phi +\frac{1}{2m} \left({\bf p} - \frac{q}{c} {\bf A} \right)^2,
\label{eq:defH} 
\end{equation} 
where  ${\bf A} $ is the vector potential, and ${\bf p}$ is the canonical momentum. Hamilton's equations imply that
\begin{equation}
\frac{d H}{d t} = q\frac{\partial \Phi}{\partial t} - \frac{q {\bf v}}{c}\cdot \frac{\partial {\bf A}}{\partial t},
\label{eq:dHdt} 
\end{equation} 
where ${\bf v} = m^{-1} ({\bf p} - q{\bf A}/c)$ is the particle's
velocity.  The electric field is given by ${\bf E} = -\nabla \Phi -
c^{-1} \partial{\bf A}/\partial t $. The second term in
equation~(\ref{eq:dHdt}) is $q{\bf v} \cdot {\bf E}_{\rm s}$, where
${\bf E}_{\rm s} = -c^{-1} \partial{\bf A}/\partial t$ is the part of
the electric field that has a nonzero curl.  In AWs and KAWs with
$\omega < \Omega_{\rm p}$ and $k_\perp \rho_{\rm p}
\lesssim 1$, ${\bf E}_{\rm s}$ is negligible compared to the total
electric field in low-$\beta $ plasmas~[see equation~(46) of
\cite{hollweg99c}], which are our primary focus, and so from here on we
neglect the second term in equation~(\ref{eq:dHdt}).

When $\partial \Phi/\partial t > 0$, a particle can gain potential
energy, kinetic energy, or both.  For example, if an ion interacts
with an electrostatic wave with wavelength $\gg \rho  $ and
frequency~$\ll \Omega  $, then the ion's guiding center drifts
with velocity $c {\bf E} \times {\bf B}_0/B_0^2$.  The particle's
kinetic energy undergoes small-amplitude oscillations due to its
gyro-motion. However, because its guiding center moves perpendicular
to $\nabla \Phi$, there is no significant secular change in its
kinetic energy.  The ion's magnetic moment~$\mu$ is almost exactly
conserved, and the change in its total-energy is almost exactly equal
to the change in its potential energy.

\begin{figure}[h]
   \centerline{\includegraphics[width=8.cm]{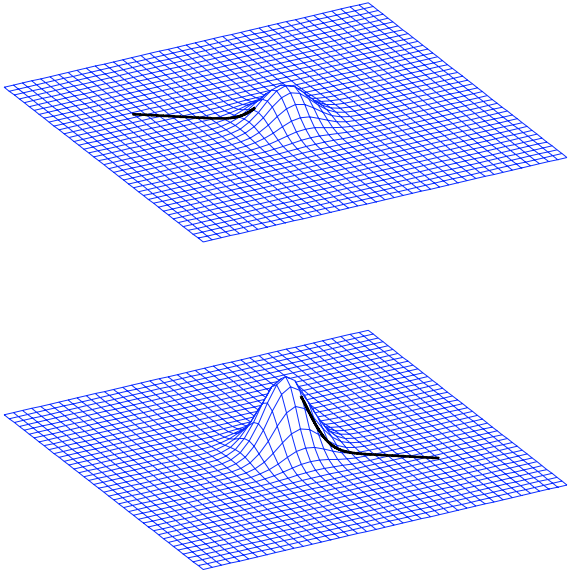}}
\caption{The potential~$\Phi$ in the~$x_1-x_2$~plane at some initial time (upper surface), at a later time (lower surface), and along the trajectory of a particle moving in a straight line in the $x_1 - x_2$~plane (thick solid line). The ``potential energy hill'' is shorter when the particle rolls to the top, and higher when the particle rolls down, so the particle gains kinetic energy as it rolls over the hill.}
\label{fig:hill1_and_hill2}
\end{figure}

On the other hand, if a particle enters a region in which $\partial
\Phi/\partial t >0$ and then leaves this region, moving up and down the
potential gradient, then it can gain kinetic energy as illustrated in
figure~\ref{fig:hill1_and_hill2}. The ``wire-mesh'' surface in the upper
panel of this figure represents $\Phi(x_1,x_2,t)$ at some initial time, and
the lower panel shows $\Phi(x_1,x_2,t)$ at a later time. We
take the maximum of $\Phi$ to be located at $\sigma=0$, where $\sigma\equiv
\sqrt{x_1^2 + x_2^2}$.  We have assumed that $\partial \Phi/\partial t >0$
at $\sigma \lesssim \sigma_0$ and $\partial \Phi/\partial t = 0$ at $\sigma \gtrsim \sigma_0$,
where $\sigma_0$ is the approximate radius in the $x_1-x_2$~plane of the
``potential-energy hills'' that appear in the figure.  The thick solid line
shows the value of $\Phi$ along the trajectory of a particle moving in a
straight line in the $x_1-x_2$~plane.  Because $\partial \Phi/\partial
t>0$, the potential-energy hill is shorter when the particle is ``climbing
up'' and higher when the particle is ``rolling down.'' The particle thus
experiences a net gain of kinetic energy from ``rolling over the hill.''

We now estimate the rate at which ions are heated by AW or KAW
fluctuations with $\lambda_\perp \sim \rho $. We note that the
condition $\lambda_\perp \sim \rho $ is intended to encompass
structures with $\lambda_\perp = 0.5 \rho $, which we invoke below
when discussing equation~(\ref{eq:omegaeff2}). However, we ignore
fluctuations with $\lambda_\perp \gg \rho$ or $\lambda_\perp \ll \rho$
throughout this discussion. Although we are interested in stochastic
ion orbits, we can still define an effective guiding-center position,
\begin{equation}
{\bf R} = {\bf r} + \frac{{\bf v} \times {\bf \hat{b}}}{\Omega  },
\label{eq:defR} 
\end{equation} 
where ${\bf \hat{b}} = {\bf B}/B$ and ${\bf r}$ is the ion's instantaneous
position. When $\epsilon \ll 1$, the particle gyrates smoothly about
position~${\bf R}$. As $\epsilon$ increases towards~1, the particle's
motion becomes more complicated, but the particle remains within a
distance~$\sim \rho $ of position~${\bf R}$.  Taking the time derivative of equation~(\ref{eq:defR}) and using the equation $d{\bf v}/dt = (q/m) ({\bf E} + {\bf v} \times {\bf B} /c)$, we obtain the equation
\begin{equation}
\frac{d{\bf R}}{dt} = v_\parallel {\bf \hat{b}} + \frac{c{\bf E} \times {\bf B}}{B^2} + \dots,
\label{eq:Rdot} 
\end{equation} 
where the ellipsis ($\dots$) represents terms proportional to derivatives of~${\bf
  B}$, which we ignore in our approximate treatment. During a single
cyclotron period, an ion passes through a small number of uncorrelated
fluctuations or ``structures'' of transverse scale~$\sim \rho $. Within different
structures,  the vector $c{\bf E}\times{\bf B}/B^2$ has a similar magnitude
($\sim \delta v_{\rho}$) but points in different directions. The time
average of $c{\bf E}\times{\bf B}/B^2$ over a single cyclotron period is thus
somewhat smaller than, but of order,~$\delta v_\rho$. The time~$\Delta t$
required for an ion's guiding center to move a distance~$\rho $ is
thus approximately 
\begin{equation}
\Delta t \sim \frac{\rho }{\delta v_\rho}.
\label{eq:dt1} 
\end{equation} 
[In writing equation~(\ref{eq:dt1}), we have assumed that the gyro-scale
fluctuations do not oscillate on a time scale $\ll \Delta t$, and we
continue to make this assumption in the analysis to follow.]  Each time the
particle moves a distance~$\rho $ perpendicular to~${\bf B}_0$, it
encounters different and uncorrelated gyro-scale electromagnetic fields. Thus,  $d{\bf R}/dt$  decorrelates after a time~$\Delta t$, and the particle's guiding center undergoes a random
walk in space with diffusion coefficient~$\sim \rho ^2/\Delta t$.

Similarly, when $\epsilon$ is sufficiently large that the ion's motion
becomes stochastic, the value of $dH/dt$ decorrelates after a
time~$\Delta t$, and the particle undergoes a random walk in energy.
In contrast, as shown in Appendix~\ref{ap:mu}, as $\epsilon
\rightarrow 0$ the interaction between ions and gyro-scale
electrostatic-potential structures is not a Markov process; instead,
changes in~$H$ are correlated over long times, and to leading order
in~$\epsilon$ are reversible and bounded.  Returning to
the stochastic case, we define $\overline{ \partial \Phi/\partial t}$
to be the rms value of $\partial \Phi/\partial t$ associated with
fluctuations with $\lambda_\perp \sim \rho $. The rms change in~$H$
during a time~$\Delta t$ is then
\begin{equation}
\Delta H \sim q\,\overline{ \frac{\partial \Phi}{\partial t}}\,\Delta t.
\label{eq:dH1} 
\end{equation} 
An ion undergoing stochastic motion can gain kinetic energy in the
same way as the particle illustrated in
figure~\ref{fig:hill1_and_hill2}. If the ion spends a time~$\Delta t$
localized within a flux tube of cross-sectional area~$\sim \rho^2$ and
length~$\sim |v_\parallel| \Delta t$, it exits this flux tube in a
random direction. Thus, if $\partial \tilde{\Phi}/\partial t$ is on average
positive during this time interval within the flux tube, it does not
follow that the ion will move to a region of larger~$\tilde{\Phi}$
after a time~$\Delta t$, where $\tilde{\Phi}$ is the electrostatic
potential associated with fluctuations with $\lambda_\perp \sim \rho
$. On the contrary, the change in~$\tilde{\Phi}$ along the ion's path
is only loosely correlated with the average change in~$\tilde{\Phi}$
within the flux tube. As a result, the change in the ion's kinetic
energy during a time~$\Delta t$ is of the same order of magnitude as
the change in its total energy given in
equation~(\ref{eq:dH1}).\footnote{In contrast, in the small-$\epsilon$
  limit addressed in Appendix~\ref{ap:mu}, the change in a particle's
  total energy is almost exactly equal to the change in the
  gyro-averaged potential energy.}  Because $\nabla \Phi$ is nearly
perpendicular to~${\bf B}$, and because the ion's guiding center moves
perpendicular to~${\bf B}$ by a distance of order $\lambda_\perp \sim
\rho$ during a time~$\Delta t$, the ion's perpendicular kinetic
energy~$K_\perp = mv_\perp^2/2$ changes by an amount of order
\begin{equation}
\Delta K_\perp \sim \Delta H
\label{eq:dK1} 
\end{equation} 
during a time~$\Delta t$. We discuss the parallel kinetic energy following equation~(\ref{eq:Q2}) below. We define an effective frequency~$\omega_{\rm eff}$ for gyro-scale fluctuations through the equation
\begin{equation}
\overline{ \frac{\partial \Phi}{\partial t}} = \omega_{\rm eff}\, \delta \Phi_\rho.
\label{eq:omegaeff} 
\end{equation} 
For example, if the gyro-scale fluctuations consist of waves with a
single frequency~$\omega$, then $\omega_{\rm eff} = \omega$. With the
use of equations~(\ref{eq:dPhi1}) and (\ref{eq:dt1}), we can rewrite
equation~(\ref{eq:dK1}) as
\begin{equation}
\Delta K_\perp \sim m v_\perp \omega_{\rm eff} \rho .
\label{eq:dK2} 
\end{equation} 
The kinetic-energy diffusion coefficient $D_K\sim (\Delta K_\perp)^2/\Delta t$ is then given by
\begin{equation}
D_K \sim m^2 v_\perp^2 \omega_{\rm eff}^2 \,\delta v_\rho \, \rho .
\label{eq:DH1} 
\end{equation} 

When a single ion undergoes kinetic-energy diffusion, the ion has an equal likelihood of gaining or losing kinetic energy during each ``random-walk step'' of duration~$\Delta t$.
On the other hand, if a large population of ions undergoes kinetic-energy diffusion, and if the phase-space density~$f$ of ions is a monotonically decreasing function of~$K_\perp$, then the average value of~$K_\perp$ increases steadily in time. To distinguish between properties of individual particles and rms quantities within a distribution, we define $v_{\perp {\rm i}}$
to be the rms perpendicular velocity of the ions, which is related to the perpendicular ion temperature~$T_\perp$ by the equation
\begin{equation}
v_{\perp {\rm i}} = \sqrt{\frac{2 k_{\rm B} T_\perp }{m}}.
\label{eq:vperpi} 
\end{equation} 
We also define the rms ion gyroradius,
\begin{equation}
\rho_{\rm i} = \frac{v_{\perp{\rm i}}}{\Omega}.
\label{eq:rhoi} 
\end{equation} 
We define $\delta v_{\rm i}$ to be the rms amplitude of the fluctuating fluid velocity at $\lambda_\perp \sim \rho_{\rm i}$, and we set
\begin{equation}
\epsilon_{\rm i}  = \frac{\delta v_{\rm i}}{v_{\perp {\rm i}}}.
\label{eq:epsi} 
\end{equation} 
For protons, we define $\delta v_{\rm p}$ ($\delta B_{\rm p}$) to be
the fluctuating fluid velocity (magnetic field) at $\lambda_\perp \sim
\rho_{\rm p}$, and we define
\begin{equation}
\epsilon_{\rm p}  = \frac{\delta v_{\rm p}}{v_{\perp {\rm p}}}.
\label{eq:epsp} 
\end{equation} 
The time scale for the average value of~$K_\perp$ in a distribution of ions to double is then roughly
\begin{equation}
t_{\rm i} \sim \frac{m^2 v_{\perp{\rm i}}^4 }{D_{K{\rm i}}},
\label{eq:th} 
\end{equation} 
where $D_{K{\rm i}}$ is the value of~$D_{\rm K}$ for ions with $v_\perp = v_{\perp {\rm i}}$
and $\rho = \rho_{\rm i}$.  The perpendicular ion heating rate per unit
mass is then $Q_\perp \sim v_{\perp{\rm i}}^2/t_{\rm i}$, or
\begin{equation}
Q_\perp \sim  \omega_{\rm eff,i}^2 \,\delta v_{\rm i} \rho_{\rm i},
\label{eq:Q1} 
\end{equation} 
where $\omega_{\rm eff,i}$ is the value of~$\omega_{\rm eff}$ at $\rho = \rho_{\rm i}$.

We now consider what determines the value of $\omega_{\rm eff}$ in
anisotropic AW or KAW turbulence. If the turbulence is driven at an
``outer scale''~$L_0$ that is $\gg \rho$, the advection or
``sweeping'' of structures with~$\lambda_\perp \sim \rho$ by
the outer-scale velocity fluctuations leads to rapid time variations
in~$\Phi$ at a fixed point in space.  On the other hand, these
large-scale velocity fluctuations advect both the ions and the
small-scale structures in the electric and magnetic fields.  Thus, if
one considers ions within a flux tube of radius~$\sim \rho$
and length~$\ll L_0$, and if one transforms to a frame of reference
moving with the average velocity of that flux tube, then the rapid time
variations resulting from large-scale advection disappear. This
indicates that large-scale sweeping does not control the rate of ion
heating or the value of $\omega_{\rm eff}$ in
equation~(\ref{eq:Q1}). On the other hand,
electrostatic-potential structures at scale~$\lambda_\perp \simeq 0.5
\rho$ are advected by
velocity fluctuations at the same scale, and there is no frame of reference in which the
velocities at $\lambda_\perp \simeq 0.5
\rho$ vanish at all points along an ion's gyro-orbit. This advection 
by velocity fluctuations with~$\lambda_\perp \simeq 0.5
\rho$ causes $\overline{ \partial \Phi/\partial t}$ to
have a value of $\sim \delta \Phi_{\rm \rho} \delta v_{\rho}/\rho$, which gives 
\begin{equation}
\omega_{\rm eff} \sim \frac{\delta v_{\rho}}{\rho},
\label{eq:omegaeff2} 
\end{equation} 
where we have neglected factors of order unity, such
as the ratio between~$\delta v_{\rho}$ and the rms amplitude of the
velocity fluctuation at $\lambda_\perp \simeq 0.5 \rho$.
Put another way, the advection of electrostatic-potential structures
at $\lambda_\perp \sim 0.5 \rho$, which are rooted in the
electron fluid, leads to a partial time derivative of~$\Phi$ that ions
can feel, and which energizes ions through the process illustrated in
figure~\ref{fig:hill1_and_hill2}. We note that in ``imbalanced'' (or
cross-helical) AW turbulence, in which the majority of the waves
propagate either parallel to~${\bf B}_0$ or anti-parallel to~${\bf
  B}_0$, the energy cascade time for the majority waves can greatly
exceed~$\omega_{\rm eff}^{-1}$, since the majority waves are cascaded
by the smaller-amplitude waves propagating in the opposite direction.  Nevertheless,
even for imbalanced turbulence, the arguments leading to
equation~(\ref{eq:omegaeff2}) continue to hold.

As discussed following equation~(\ref{eq:defeps}) and in
Appendix~\ref{ap:mu}, when $\epsilon$ is sufficiently small, the
changes in~$H$ remain correlated (and largely reversible) over long
times, so that the perpendicular heating rate is strongly reduced
relative to our estimate in equation~(\ref{eq:Q1}). To account for
this, we introduce a multiplicative suppression factor onto the
right-hand side of equation~(\ref{eq:Q1}) of the form~$\exp
(-c_2/\epsilon_{\rm i})$.  We also add an overall coefficient~$c_1$ to the
right-hand side of equation~(\ref{eq:Q1}) to account for the various
approximations we have made.  Both $c_1$ and $c_2$ are dimensionless
constants whose values depend upon the nature of the fluctuations
(e.g., whether the fluctuations are waves or turbulence, the type of
turbulence, etc) and the shape of the ion velocity distribution.
Substituting equation~(\ref{eq:omegaeff2}) into
equation~(\ref{eq:Q1}), we obtain
\begin{equation}
Q_\perp = \frac{c_1(\delta v_{\rm i})^3}{\rho_{\rm i}}
\,\exp\left(-\,\frac{c_2}{\epsilon_{\rm i}}\right).
\label{eq:Q2} 
\end{equation} 
We emphasize that for protons in low-$\beta$ plasmas, $\epsilon_{\rm p} \simeq
\beta^{-1/2} \delta B_{\rm p}/B_0$, and thus $\epsilon_{\rm p}$ can approach
unity even if $\delta B_{\rm p}/B_0$ remains~$\ll 1$.

The change in an ion's parallel kinetic energy $K_\parallel =
mv_\parallel^2/2$ during a time~$\Delta t$ due to the parallel
electric field~$E_\parallel$ is $\Delta K_\parallel \sim qE_\parallel
v_\parallel \Delta t$.  We have restricted our analysis to AWs and
KAWs with $\lambda_\perp \sim \rho \gtrsim \rho_{\rm p}$ and
$\omega^{-1} \gtrsim \Delta t \sim \lambda_\perp/\delta v_\rho$.  This
condition on the wave frequency implies that the parallel wavelengths
of such fluctuations satisfy the inequality $\lambda_\parallel\gtrsim
\rho v_{\rm A}/\delta v_\rho \gg \lambda_\perp$.  When $m_e/m_p <
\beta < 1$ and $\lambda_\perp > \rho_{\rm p}$, $E_\parallel / E_\perp
\sim \rho_{\rm p}^2/(\lambda_\perp \lambda_\parallel)$, where $m_e$ is
the electron mass~\citep{hollweg99c} and $E_\perp$ is the
electric-field component perpendicular to~${\bf B}$. Thus $\Delta
K_\parallel$ is $\lesssim v_\parallel/v_{\rm A}$ times the value of
$\Delta K_\perp$ in equation~(\ref{eq:dK2}). For thermal ions in
low-$\beta$ plasmas, $v_\parallel \ll v_{\rm A}$.  Thus, when
$\epsilon_{\rm i}$ is sufficiently large that stochastic heating is
important, stochastic heating leads primarily to perpendicular ion
heating rather than parallel heating.  For AWs/KAWs in low-$\beta$
plasmas, the parallel component of the magnetic mirror force is much
less than $qE_\parallel$ \citep{hollweg99c} and thus does not affect
our conclusions regarding anisotropic heating at $\beta \ll 1$.

\subsection{Test-Particle Simulations of Proton Heating}
\label{sec:tp} 

To test the above ideas, we have numerically simulated test-particle
protons interacting with a spectrum of randomly phased KAWs. The
protons' initial locations are chosen randomly from a uniform
distribution within a volume encompassing many wavelengths
perpendicular and parallel to~${\bf B}_0$. The protons' initial
velocities are drawn randomly from an isotropic Maxwellian
distribution of temperature~$T_{\rm p}$. For each particle, we solve
the equations
\begin{equation}
\frac{d {\bf x} }{dt} = {\bf v} 
\label{eq:dxdt} 
\end{equation} 
and 
\begin{equation}
\frac{d{\bf v}}{dt} = \frac{q}{m}\left({\bf E} + \frac{{\bf v} \times {\bf B} }{c}\right)
\label{eq:dvdt} 
\end{equation} 
using the Bulirsch-Stoer method~\citep{press92}. We take ${\bf B} =
B_0 {\bf \hat{z}} + {\bf B}_1$, where $B_0$ is constant. We take ${\bf
  E}$ and ${\bf B}_1$ to be the sum of the electric and magnetic
fields from 162 waves with randomly chosen initial phases, with two
waves at each of 81 different wave vectors. At each wave vector, there
is one wave with $\omega/k_z > 0$ and a second wave with $\omega/k_z <
0$.  This second wave has the same amplitude as the first, so that
there are equal fluxes of waves propagating in the $+z$ and $-z$
directions. The 81 different wave vectors consist of 9 wave vectors at
each of nine different values of~$k_\perp$, denoted $k_{\perp j}$. The
$k_{\perp j}$ can be expressed in terms~$\rho_{\rm p}$. In particular,
the values $\psi_j = \ln (k_{\perp j} \rho_{\rm p})$ are uniformly
spaced between $-4/3$ and $4/3$; i.e., $\psi_j = -4/3 + j/3$, with $j=
0, 1, \dots , 8$. We regard the values $\psi_j$ as corresponding to
cell centers in a uniform grid in $\psi = \ln (k_\perp \rho_{\rm p}
)$, with grid spacing $\Delta \psi = 1/3$. The middle three grid
cells, with $j = \mbox{3, 4, and 5}$, thus correspond to an interval
of width unity in $\ln(k_\perp)$ space centered on $k_\perp \rho_{\rm
  p} = 1$.  We define the rms amplitudes of the gyro-scale velocity
and magnetic-field fluctuations $\delta v_{\rm p}$ and $\delta B_{\rm
  p}$ in our simulations by taking the rms values of the ${\bf
  E}\times {\bf B}$ velocity and magnetic-field fluctuation resulting
from the KAWs in these middle three grid cells.  At each $k_{\perp j}$
we include 9 different values of the azimuthal angle~$\phi$ in $k$
space, $\phi_l = 2\pi l/9$, where $l = 0, 1, \dots , 8$. At each
$k_{\perp j}$ there is only a single value of~$k_\parallel$, which we
denote $k_{\parallel j}$.  We choose $k_{\parallel 4}$ so that the
frequency at $k_\perp = k_{\perp 4}$ and $k_\parallel = k_{\parallel
  4}$ equals $k_{\perp 4} \delta v_\rho$. The linear frequency of our
gyro-scale KAWs is thus comparable to the value of~$\omega_{\rm eff}$
given in equation~(\ref{eq:omegaeff2}) for KAW turbulence at $k_\perp
\rho_{\rm p} \sim 1$. We then set
\begin{equation}
\frac{k_{\parallel j}}{k_{\parallel 4}} = \left\{ 
\begin{array}{ll}
 (k_{\perp j}/k_{\perp 4})^{2/3} & \mbox{ if $0 \leq j<4$} \\
 (k_{\perp j}/k_{\perp 4})^{1/3} & \mbox{ if $4 < j \leq 8$} 
\end{array}\;.
\right.
\label{eq:kpar1} 
\end{equation} 
Our formula for $k_{\parallel j}$ at $j< 4$ is chosen so that the wave
periods are comparable to the energy cascade time scales in the
critical-balance theory of \cite{goldreich95}, while the formula for
$j>4$ is chosen so that the wave periods match the nonlinear time
scales in the critical-balance theory of~\cite{cho04b}.  All waves at
the same~$k_\perp$ have the same amplitude, and (since there are the
same number of waves at each~$k_{\perp j}$) we take the amplitude of
the magnetic-field fluctuation in each wave to be~$\propto
k_\perp^{-1/3}$ for $k_\perp \rho_{\rm p} < 1$ and $\propto
k_\perp^{-2/3}$ for $k_\perp \rho_{\rm p} > 1$, again motivated by
the theories of \cite{goldreich95} and \cite{cho04b}. 

The relative amplitudes of the different components of ${\bf E}$ and
${\bf B}_1$ for each wave are taken from the two-fluid theory of
\cite{hollweg99c}. To apply this theory, we choose plasma parameters
that are characteristic of coronal holes. In particular, we set
$\beta_e = 8\pi n k_{\rm B} T_{\rm e}/B_0^2 = 0.003$, $v_{\rm A} =
0.003c$, and $T_{\rm e} = 0.5 T_{\rm p}$, where $n$ is the electron
number density (equal to the proton number density), and $T_{\rm e}$
is the electron temperature.

Using the above procedures, we have carried out seven simulations with
different values for the overall normalization of the wave amplitudes,
with $\delta B_{\rm p}/B_0$ ranging from $4.8\times 10^{-3}$ to~$1.9
\times 10^{-2}$.  Given the polarization properties of KAWs and our
method for constructing the wave spectra, the value of $\delta v_{\rm
  p}/v_{\rm A}$ is $1.19$ times the value of~$\delta B_{\rm p}/B_0$ in
each simulation. The wave frequencies reach their maximum values in
the largest-$\delta B_{\rm p}/B_0$ simulation.  In this simulation,
$\omega = 0.29\Omega_{\rm p}$ at $k_\perp \rho_{\rm p } = 1$, and
$\omega = 0.82 \Omega_{\rm p}$ at the maximum value of $k_\perp
\rho_{\rm p}$, which is~$ 3.79$.  Although this maximum frequency is
close to~$\Omega_{\rm p}$, the cyclotron resonance condition $\omega -
k_\parallel v_\parallel = l \Omega_{\rm p}$ (where $l$ is any integer)
is not satisfied, because the parallel thermal speed of the protons is
only~$0.055 v_{\rm A}$ and $k_\parallel \ll k_\perp$. For most of the
waves in these simulations,~$\omega \ll \Omega_{\rm p}$.

\begin{figure}[t]
\centerline{\includegraphics[width=8.cm]{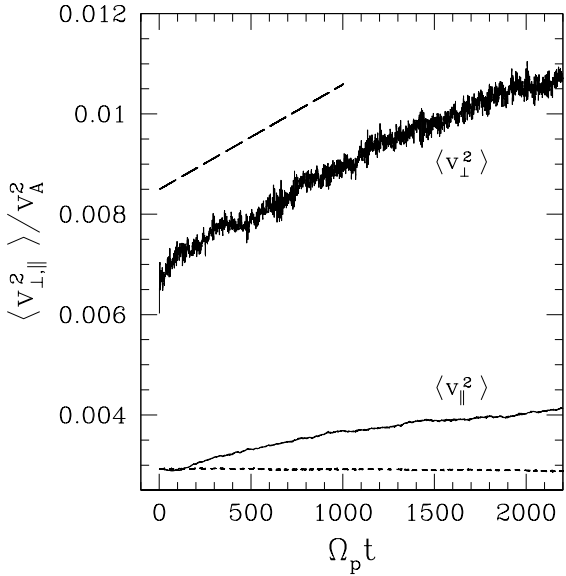}}
\caption{$v_\perp^2$ and $v_\parallel^2$ averaged over the $10^3$ particles in a simulation with $\delta v_{\rm p}/v_{\perp{\rm p}} = 0.15$, $\beta_e = 0.003$, $v_{\rm A} =
0.003c$, and $T_{\rm e} = 0.5 T_{\rm p}$. The two solid-line curves correspond to our basic numerical method.  We determine $Q_{\perp {\rm p}}$ in this simulation from the slope of the long-dashed line. The short-dashed line shows $\langle v_\parallel^2 \rangle$ in a modified simulation with the same parameters in which ${\bf E}$ is replaced by  ${\bf E}^\prime = 
{\bf E} + {\bf \hat{b}} (E_z
- {\bf \hat{b}} \cdot {\bf E})$.
\label{fig:vperppar}}\vspace{0.5cm}
\end{figure}

We determine the perpendicular proton heating rate per unit mass~$Q_{\perp {\rm p}}$ in the
simulations by plotting $\langle v_\perp^2\rangle$ versus time,
fitting this plot to a straight line to determine $(d/dt) \langle
v_\perp^2\rangle$, and then setting $Q_{\perp {\rm p}} = 0.5(d/dt) \langle
v_\perp^2\rangle$, where $\langle \dots \rangle$ indicates an average
over the $10^3$ particles in each simulation. When fitting the plot of
$\langle v_\perp^2\rangle$ versus time, we ignore the first $10$
cyclotron periods, because during the first couple gyro-periods the
particles undergo a modest apparent heating as they ``pick up'' some
portion of the ${\bf E} \times {\bf B} $ velocity of the waves. We
find that after $\langle v_\perp^2\rangle$ increases by between 20\%
and 40\%, the heating rate starts
to decrease for two reasons. First,  the small-$v_\perp$ part of the
velocity distribution flattens, after which this part of the
distribution is no longer heated as effectively. Second, 
as $\langle v_\perp^2\rangle$ increases,
$\epsilon_{\rm p}$ decreases.  We neglect this later stage of weaker heating when
constructing our fits to the $\langle v_\perp^2(t) \rangle$ plots, so
that the measured heating rates correspond to Maxwellian
distributions. (For the smallest values of $\delta v_{\rm p}$, we do not
observe a second stage of weaker heating, because the test-particle
velocity distributions do not change very much during the simulations,
which last $10^4 \Omega_{\rm p}^{-1}$.)  We illustrate this procedure
in figure~\ref{fig:vperppar} for a run with $\delta
v_{\rm p}/v_{\perp {\rm p}} = 0.15$.  In this case, we determine
$Q_{\perp {\rm p}}$ from the slope of the long-dashed line, which is our fit to the
$\langle v_\perp^2 \rangle$ data over the interval $10 < \Omega_{\rm
  p} t < 10^3$.

\begin{figure}[t]
\centerline{\includegraphics[width=8.cm]{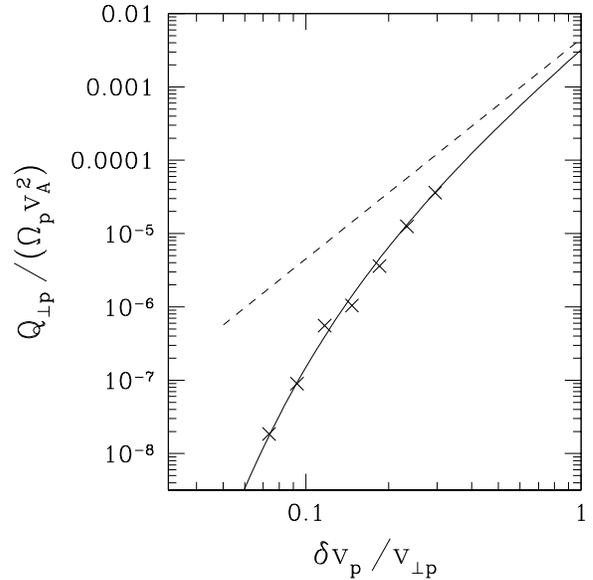}}
\caption{Numerical results (plotted with $\times$s) for the perpendicular heating rate~$Q_{\perp {\rm p}}$ for test-particle protons interacting with a spectrum of randomly phased KAWs. The solid line is equation~(\ref{eq:Q3}), and the dashed line is equation~(\ref{eq:Q3}) with the 
``$\mu$ conservation'' factor~$\exp(-0.34/\epsilon_{\rm p})$ replaced with unity.}\vspace{0.5cm}
\label{fig:Q1}
\end{figure}

In figure~\ref{fig:Q1} we plot the values of~$Q_{\perp {\rm p}}$ for several
different values of~$\epsilon_{\rm p}$.
Each $\times$ in this figure corresponds to a separate simulation with a different
value of $\delta v_{\rm p}$ but the same initial proton temperature.  The solid line is
the proton heating rate from equation~(\ref{eq:Q2}) with $c_1 = 0.75$
and $c_2 = 0.34$; that is,
\begin{equation}
Q_{\perp {\rm p}} = \frac{0.75 (\delta v_{\rm p})^3}{\rho_{\rm p}}
\,\exp\left(-\,\frac{0.34}{\epsilon_{\rm p}}\right).
\label{eq:Q3} 
\end{equation} 
We expect the constants $c_1$ and $c_2$ to be fairly insensitive to
variations in $\beta_{\rm e}$, $T_{\rm p}/T_{\rm e}$, and $v_{\rm
  A}/c$ (at least within the range of solar-wind-relevant parameters),
in which case $Q_{\perp {\rm p}}$ depends on the plasma parameters
primarily through the explicit $\rho_{\rm p}$ and $\epsilon_{\rm p}$
terms in equation~(\ref{eq:Q3}).  The values of $c_1$ and $c_2 $ in
equation~(\ref{eq:Q3}) presuppose the presence of a broad spectrum of
AWs and KAWs bracketing the perpendicular wavenumber $k_\perp =
(\rho_{\rm p})^{-1}$, encompassing at a minimum the range $0.3
\lesssim k_\perp \rho_{\rm p} \lesssim 3$. A spectrum of at least this
width is probably present in the solar wind, the only uncertainty
being the value of the dissipation wavenumber beyond which the wave
power spectrum decreases exponentially with increasing~$k_\perp$. If
the simulations described in this section are repeated without the
smallest three values of~$k_\perp$ and without the largest three
values of $k_\perp$ (keeping the wave amplitudes fixed at the middle
three values of $k_\perp$), then the proton orbits become less
stochastic, and $Q_{\perp {\rm p}}$ decreases significantly. (The
exact amount by which $Q_{\perp {\rm p}}$ decreases depends upon the
value of~$\epsilon_{\rm p}$.) We have omitted waves at $k_\perp
\rho_{\rm p} < 0.26$ and $k_\perp \rho_{\rm p} > 3.8$, but we expect
that waves at such scales do not have a strong effect on perpendicular
ion heating, provided $\omega$ is sufficiently small that the
cyclotron resonance condition can not be satisfied. It is possible
that in some cases strongly turbulent fluctuations with $k_\perp
\rho_{\rm p} \gg 1$ and nonlinear time scales $\sim \Omega_{\rm
  p}^{-1}$ could heat ions through a broadened cyclotron resonance,
but a detailed investigation of this process is beyond the scope of
this study.

We reiterate that the values of $c_1$ and $c_2$ in
equation~(\ref{eq:Q3}) are not universal, but instead depend on the
type of fluctuations that are present. In true turbulence (as opposed
to random-phased waves), the value of $c_2$ may be smaller than in our
simulations (indicating stronger heating), because a significant
fraction of the cascade power may be dissipated in coherent structures
in which the fluctuating fields are larger than their rms
values~\citep{dmitruk04}.

The lower solid-line curve in figure~\ref{fig:vperppar} plots $\langle
v_\parallel^2 \rangle$ versus time in the simulation with
$\epsilon_{\rm p} = 0.15$, $\beta_e = 0.003$, $v_{\rm A} = 0.003c$,
and $T_{\rm e} = 0.5 T_{\rm p}$.  During the interval $10 < \Omega_{\rm
  p}t < 2200$, the increase in $\langle v_\parallel^2 \rangle$ is
about one-fourth the increase in $\langle v_\perp^2 \rangle$. However,
most of the increase in $\langle v_\parallel^2 \rangle$ is an artifact
of our numerical method, which equates the parallel electric fields of
the waves with the $z$ component of the electric field in the
simulation, and the perpendicular electric field of the waves with the
$x$ and $y$ components of the electric field in the simulation. The
local magnetic field in our simulations, however, is not parallel to
the $z$ axis, but instead has nonzero $x$ and~$y$ components resulting
from the magnetic-field fluctuations.  As a result, part of the
perpendicular wave electric field is converted into a parallel
electric field in the simulation, artificially enhancing the parallel
electric field seen by the particles. To eliminate this effect, we
have repeated this simulation replacing the local electric
field~${\bf E}$ seen by each particle with the adjusted electric
field~${\bf E}^\prime = {\bf E} + {\bf \hat{b}} (E_z - {\bf \hat{b}}
\cdot {\bf E})$, where ${\bf \hat{b}}= {\bf B}/B$ and ${\bf B}$ is the
local value of the magnetic field.  In this new simulation, the
parallel electric field ${\bf \hat{b}} \cdot {\bf E}^\prime$ is the
sum of the parallel electric fields of the individual waves in the
simulation and does not include any contribution from the
perpendicular electric fields of the individual waves. The value of
$\langle v_\parallel^2 \rangle$ in this modified simulation, shown as
a dashed line in figure~\ref{fig:vperppar}, does not increase
significantly during the course of the simulation (in fact it
decreases slightly), consistent with our argument above that parallel heating
is weak when~$\beta \ll 1$.

\subsection{Proton Heating at $k_\perp \rho_{\rm p} \sim
  1$ as a Fraction of the Turbulent Cascade Power}
\label{sec:cascade} 

The cascade power per unit mass at $k_\perp \rho_{\rm p} \sim 1$,
which we denote~$\Gamma$, depends upon whether the turbulence is
``balanced'' or ``imbalanced,'' where balanced (imbalanced) turbulence
involves equal (unequal) fluxes of waves propagating parallel to~${\bf
  B}_0$ and anti-parallel to~${\bf B}_0$.  In balanced KAW turbulence,
\begin{equation}
\Gamma = C_{\rm K}^{-3/2} \left(\frac{ \delta v_{\rm p}}{\rho_{\rm p}}\right) \left(\frac{\delta B_{\rm p}}{B_0}\right)^2 v_{\rm A}^2,
\label{eq:Gamma1} 
\end{equation} 
where $C_{\rm K}$ is a dimensionless constant (Howes et
al. 2008a)\nocite{howes08a}.  It can be inferred from the numerical
simulations of Howes et al.\ (2008b)\nocite{howes08b} that~$C_{\rm K} = 2.0$
(G. Howes, private communication). In the simulations of
section~\ref{sec:tp}, $\delta B_{\rm p} /B_0 = 0.84 \delta v_{\rm p}/ v_{\rm
  A}$, and we make the approximation that this same ratio is characteristic
of KAW turbulence in general.  Combining equations~(\ref{eq:Q3}) and
(\ref{eq:Gamma1}), we obtain
\begin{equation}
\frac{Q_{\perp{\rm p}}  }{\Gamma} = 3.0 \,\exp\left(-\,\frac{0.34}{\epsilon_{\rm p}}\right).
\label{eq:QG} 
\end{equation} 
We expect that~$C_K$, like the constants~$c_1$ and $c_2$, depends only
weakly on~$\beta$, $T_{\rm p}/T_{\rm e}$, and $v_{\rm A}/c$ (at least
for solar-wind-relevant parameters), so that the numerical constants
3.0 and 3.4 in equation~(\ref{eq:QG}) are relatively insensitive to the
plasma parameters.  Equation~(\ref{eq:QG}) implies that perpendicular
proton heating by KAWs with $k_\perp \rho_{\rm p } \sim 1$ absorbs
$\geq 1/2$ of the cascade power at $k_\perp \rho_{\rm p} \sim 1$ when
$\epsilon_{\rm p}$ exceeds
\begin{equation}
\epsilon_{\rm crit} = 0.19 \; .
\end{equation} 

The cascade power in imbalanced AW turbulence is smaller than in
balanced AW turbulence with the same total fluctuation energy, because
the AW energy cascade requires interactions between oppositely
propagating waves~\citep{iroshnikov63, kraichnan65}. At $k_\perp
\rho_{\rm p} \sim 1$, KAWs propagating in the same
direction can interact nonlinearly with one another, but the
importance of such interactions relative to interactions between
oppositely propagating waves is not well known. Despite this
uncertainty, we expect that if AW/KAW turbulence is imbalanced at
$k_\perp \rho_{\rm p} \sim 1$, then the cascade power at
$k_\perp \rho_{\rm p} \sim 1$ is less than in
equation~(\ref{eq:Gamma1}). On the other hand, it is unlikely that
imbalance strongly affects~$Q_{\perp {\rm p}}$ if $\delta v_{\rm p}$ is
held fixed (except for particles with $v_\parallel \sim  \pm v_{\rm A}$,
as discussed in section~\ref{sec:add}).  We thus expect perpendicular
proton heating to absorb at least
50\% of the cascade power at $k_\perp \rho_{\rm p} \sim 1$ in imbalanced turbulence even when $\epsilon_{\rm p}$ is somewhat smaller than~$0.19$.

\subsection{Proton Heating versus Electron Heating by KAWs with $k_\perp \rho_{\rm p} \sim 1$}
\label{sec:pe} 

Stochastic proton heating removes energy from KAW fluctuations with
$k_\perp \rho_{\rm p} \sim 1$, resulting in an effective
damping rate for these fluctuations, which we denote~$\gamma_{\rm
  p}$. The value of $\gamma_{\rm p}$ is given by the relation
\begin{equation}
2 \gamma_{\rm p} {\cal E}_{\rm w} = Q_{\perp {\rm p}},
\label{eq:gammap} 
\end{equation} 
where ${\cal E}_{\rm w}$ is the energy per unit mass of the KAW
fluctuations at $k_\perp \rho_{\rm p} \sim 1$.  The factor
of~2 in equation~(\ref{eq:gammap}) is included to make $\gamma_{\rm p}$
analogous to a linear wave damping rate, in the sense that the rate at
which linear waves lose energy is twice the product of the damping
rate and the wave energy. To estimate the value of $\gamma_{\rm p}$ in
AW/KAW turbulence, we use the test-particle calculations in
section~\ref{sec:tp} for a spectrum of randomly phased KAWs. We take
${\cal E}_{\rm w}$ to be the energy per unit mass of the full spectrum
of waves in these simulations. (This choice leads to a conservative
estimate of~$\gamma_{\rm p}$, since the damping is likely concentrated
in the subset of the waves with $k_\perp \rho_{\rm p}
\gtrsim 1$.) On the other hand, we continue to define $(\delta
v_{\rm p})^2$ as the mean-square ${\bf E} \times {\bf B}$
velocity associated with KAWs with values of $k_\perp$ lying within a
logarithmic interval of width unity centered on $k_\perp \rho_{\rm p} = 1$. With these definitions, ${\cal E}_{\rm w} = 2.1
(\delta v_{\rm p})^2$ in all of the simulations in
section~\ref{sec:tp}. Combining equations~(\ref{eq:Q3}) and (\ref{eq:gammap}), we obtain
\begin{equation}
\gamma_{\rm p } = 0.18 \epsilon_{\rm p} \Omega_{\rm p} \,\exp\left(-\,\frac{0.34}{\epsilon_{\rm p}}\right).
\label{eq:gammap2} 
\end{equation}

In low-$\beta$ plasmas, small-amplitude KAWs with $k_\perp \rho_{\rm
  p} = 1$ and $\omega \ll \Omega_{\rm p}$ undergo electron Landau
damping but negligible linear proton
damping~\citep{quataert98,gruzinov98,gary04a}.  Using the numerical
method described by~\cite{quataert98} and Howes et al.\ (2008a)\nocite{howes08a}, we
numerically solve the full hot-plasma dispersion relation to find the
electron damping rate~$\gamma_{\rm e}$ of KAWs with $k_\perp \rho_{\rm
  p} = 1$ and $\omega \ll \Omega_{\rm p}$ for a range of values of
$k_\parallel$, $T_{\rm p}/T_{\rm e}$, $\beta_{\rm p}$, and $v_{\rm
  A}/c$, where $\beta_{\rm p} = 8 \pi n k_{\rm B} T_{\rm p}/B_0^2$.
We find that if $m_{\rm e}/m_{\rm p} \ll \beta_{\rm e}  \ll 1$, $v_{\rm A} \ll c$, 
and $0.1 \lesssim T_{\rm p}/T_{\rm e} \lesssim 10$, then the damping
rate at $k_\perp \rho_{\rm p} = 1$ is well fit by the formula
$\gamma_{\rm e} = 9.5 \times 10^{-3}(T_{\rm e}/ T_{\rm p})^{1/2} \beta_{\rm p}^{-1/2}|k_\parallel v_{\rm A}|$, or equivalently
\begin{equation}
\gamma_{\rm e} = 9.5 \times 10^{-3}\epsilon_{\rm p} \chi^{-1} 
\left(\frac{ T_{\rm e}}{\beta_{\rm p}T_{\rm p}}\right)^{1/2}
 \Omega_{\rm p} ,
\label{eq:gammae} 
\end{equation} 
where $\chi \equiv k_\perp \delta v_{\rm p}/|k_\parallel v_{\rm A}|$.
In some theories of strong MHD turbulence $\chi\sim
1$~\citep{goldreich95,boldyrev06}. This condition, some times referred
to as critical balance, may characterize AW/KAW fluctuations in
coronal holes and the solar wind at $k_\perp \rho_{\rm p}
\sim 1$. On the other hand, if the frequencies of the waves launched
by photospheric motions are sufficiently small, then AW/KAW turbulence
at a heliocentric distance of a few solar radii may be more
``two-dimensional'' than in critical-balance models, with smaller
values of $k_\parallel$ and a larger value of $\chi$.

Combining equations~(\ref{eq:gammap2}) and (\ref{eq:gammae}), we obtain
\begin{equation}
\frac{\gamma_{\rm p}}{\gamma_{\rm e}} = 19\, \chi \,
\left(\frac{\beta_{\rm p}T_{\rm p}}{ T_{\rm e}}\right)^{1/2} 
\exp\left(-\,\frac{0.34}{\epsilon_{\rm p}}\right).
\label{eq:gammaratio} 
\end{equation} 
The ratio $\gamma_{\rm p}/\gamma_{\rm e}$ approximates the ratio of
the proton heating rate to the electron heating rate resulting from
KAW fluctuations at $k_\perp \rho_{\rm p} \sim 1$ in the low-$\beta$
conditions present in coronal holes and the near-Sun solar wind. (At
$\beta \gtrsim 1$, linear KAW damping on the protons becomes
important, increasing the proton heating rate.) We note that if the
damping time scales $\gamma_{\rm p}^{-1}$ and $\gamma_{\rm e}^{-1}$ are both
much longer than the energy cascade time at $k_\perp \rho_{\rm p} \sim
1$, then most of the fluctuation energy will cascade past the
proton-gyroradius scale to smaller scales. In that case, the division
of the turbulent heating between protons and electrons will depend
primarily upon how fluctuations dissipate at $k_\perp \rho_{\rm p} \gg
1$.

\subsection{How the Heating Rate Depends on
 $q$, $m$, $\beta$,  and $v_\parallel/v_{\rm A}$}
\label{sec:add} 

If we re-run our simulations, keeping only waves with $\omega/k_z >
0$, and consider a thermal distribution of test-particle protons with a
nonzero average velocity equal to $v_{\rm A} \hat{z}$, then the
perpendicular heating rate is strongly reduced. This is because the electric field
of an Alfv\'en wave (or KAW with $\lambda_\perp \sim \rho_p$) vanishes
(or is strongly reduced) in a reference frame moving at speed~$v_{\rm
  A}$ in the same direction as the wave along the background magnetic
field.  This effect may explain the observation that the perpendicular heating of
$\alpha$ particles in the solar wind is reduced when the differential
flow velocity of $\alpha$ particles relative to protons (in the
anti-Sunward direction) approaches~$v_{\rm A}$~\citep{kasper08}, at
least in regions where anti-Sunward propagating KAWs dominate over
Sunward-propagating KAWs. 

If we hold $\delta B_{\rm p}/B_0$ fixed
but increase~$\beta_{\rm p}$ to~1, then the perpendicular proton heating rate is
dramatically reduced, because $\epsilon_{\rm p} = \delta v_{\rm p}/v_\perp \sim
\beta^{-1/2} \delta B_{\rm p}/B_0$ decreases by a large factor. On the other hand,
the protons in these $\beta_{\rm p} \sim 1$ simulations undergo
significant parallel heating, consistent with results from linear
theory \citep{quataert98} and recent test-particle simulations of ions
propagating in numerically simulated MHD turbulence~\citep{lehe09}. 

If we re-run our simulations but use ${\rm O}^{+5}$ ions instead of
protons (but with the same temperature as the protons), then the
perpendicular heating rate is much larger. This is in large part because $\epsilon$ is 
larger for ${\rm O}^{+5}$ (and other heavy ions) than for protons at the same temperature, a point
to which we return in section~\ref{sec:sw}. Another reason for enhanced heavy-ion heating can be seen from equation~(\ref{eq:th}). We rewrite this equation with the aid of equation~(\ref{eq:omegaeff2}),  increasing $t_{\rm i}$ by $\exp(c_2/\epsilon_{\rm i})$ for the same reasons that we reduced~$Q_\perp$ by this same factor in equation~(\ref{eq:Q2}), to obtain
\begin{equation}
t_{\rm i} \sim \frac{v_{\perp{\rm i}}^2 \rho_{\rm i}}{(\delta v_{\rm i})^3} \, \exp\left(\frac{c_2}{\epsilon_{\rm i}}\right).
\label{eq:th2} 
\end{equation} 
In a number of theories of MHD turbulence, the ratio $(\delta
v_{\rho{\rm i}})^3/\rho_{\rm i}$ is relatively (or completely) insensitive to
the value of $\rho_{\rm i}$, provided $\rho_{\rm i}$ is in the
inertial range of the turbulence. On the other hand, for ion species
at equal temperatures,  $v_{\perp{\rm i}}^2$ is inversely
proportional to the ion mass. Thus, even aside from the exponential
factor in equation~(\ref{eq:th2}), the heating time scale is shorter
for heavier ions than for lighter ions at the same
temperature. 

Finally, if we repeat the simulations of section~\ref{sec:tp} for
test-particle ions with $\rho_{\rm i} \gg \rho_{\rm p}$,
and with values of $k_\perp \rho_{\rm i}$ centered on~1 so that the
gyro-scale fluctuations are now AWs, we recover similar values for the
perpendicular heating rate per unit mass. Stochastic perpendicular ion heating thus
does not require the particular polarization properties of KAWs, but
operates for both KAWs and AWs, as we have argued in our heuristic
derivation of equation~(\ref{eq:Q2}).

\subsection{Lack of Perpendicular Heating by AWs with $k_\perp \rho_{\rm i } \ll 1$}
\label{sec:lack} 

In turbulent flows, the rms variation in the velocity across a
perpendicular scale~$\lambda_\perp$, denoted $\delta
v_{\lambda_\perp}$, typically increases as some positive power
of~$\lambda_\perp$ when $\lambda_\perp$ is in the inertial range. As a
result, the variation in the electrostatic potential across an ion's
gyro-orbit is dominated by the fluctuations at the large-scale end of
the inertial range, suggesting that these large-scale fluctuations
might make an important contribution to the perpendicular heating
rate. This suggestion, however, is incorrect, because AWs with
$k_\perp \rho  \ll 1$ cause an ion's guiding center to drift
smoothly at velocity $c{\bf E} \times {\bf B}/B^2$, but do not cause
an ion's motion to become chaotic.  If one transforms to a reference
frame that moves at the velocity $c{\bf E} \times {\bf B}/B^2$
evaluated at the ion's guiding-center position, then the variation in
$q \Phi$ across the ion's gyroradius is a small fraction of $m v_\perp ^2/2$. The
ion's trajectory in the plane perpendicular to~${\bf B}_0$ in this
frame is approximately a closed circle, and the ion's magnetic
moment~$\mu$ is then nearly conserved~\citep{kruskal62}.

\section{Perpendicular Ion Heating in Coronal Holes and the Fast Solar Wind}
\label{sec:sw} 

As shown in the previous section, the  stochastic ion heating rate is a strongly increasing function of  $\epsilon_{\rm i} = \delta v_{\rm i}/v_{\perp \rm i}$. For fixed turbulence properties, the value of $\epsilon_{\rm i}$ depends upon the ion charge~$q =Ze$, the ion mass~$m = Am_{\rm p}$, and the perpendicular ion temperature~$T_\perp$. For example, if we take the rms amplitude of the turbulent velocity fluctuation at transverse scale~$\lambda_\perp $ to be given by
\begin{equation}
\delta v_{\lambda_\perp} = \alpha v_{\rm A} \left(\frac{\lambda_\perp}{L_0}\right)^a
\label{eq:dvl1} 
\end{equation} 
for $\rho_{\rm p} < \lambda_\perp < L_0$,
where $\alpha$ and $a$ are dimensionless constants and $L_0$ is the outer scale or driving scale of the turbulence, then
\begin{equation}
\epsilon_{\rm i} = \alpha \left(\frac{T_{\rm p}}{T_\perp \mu_{\rm p} \beta_{\rm p}}\right)^{(1-a)/2}
\frac{A^{(1+a)/2}}{Z^a}\left( \frac{d_{\rm p}}{L_0}\right)^a,
\label{eq:eps2} 
\end{equation} 
where $d_{\rm p } = v_{\rm A}/\Omega_{\rm p}$ is the proton inertial
length, $\beta_{\rm p } = 8\pi n_{\rm p} k_{\rm B} T_{\rm p}/B^2$,
$n_{\rm p}$ is the proton density, $T_{\rm p}$ is the proton
temperature, and $\mu_{\rm p}$ is the mean molecular weight per
proton; that is, the mass density is $\mu_{\rm p} n_{\rm p} m_{\rm
  p}$, and the Alfv\'en speed is $B/\sqrt{4\pi \mu_{\rm p} n_{\rm p}
  m_{\rm p}}$.  If the velocity power spectrum $P_k^{(v)}$ is
$\propto k_\perp^{-c_3}$ for $L_0^{-1} < k_\perp < \rho_{\rm i}^{-1}$, then 
\begin{equation}
a = \frac{c_3 - 1}{2}
\label{eq:defc3}.  
\end{equation} 

To investigate the possible role of stochastic ion heating in coronal holes and the fast solar wind, we evaluate equation~(\ref{eq:eps2}) as a function of heliocentric distance~$r$ using observationally constrained profiles for the density, temperature, and field strength. We take $n_{\rm p}$ to be given by equation~(4) of \cite{feldman97},
which describes coronal holes out to several solar radii, plus an
additional component proportional to~$r^{-2}$:
\begin{equation}
n_{\rm p}(r) = \left(\frac{3.23 \times 10^8}{d^{15.6}} + \frac{2.51 \times 10^6}{d^{3.76}} + \frac{1.85 \times 10^5}{d^2}\right) \mbox{ cm}^{-3},
\label{eq:n} 
\end{equation} 
where $d = r/R_{\sun}$.  Equation~(\ref{eq:n}) gives $n_{\rm p}=4
\mbox{ cm}^{-3}$ at 1~AU.  We set
\begin{equation}
T_{\rm p } = 3\times 10^6 \mbox{ K} \cdot \left[\frac{ 1 - (2/3) \exp(-d/1.5)}{(1 + 0.1d)^{0.8}}\right],
\label{eq:Tp1} 
\end{equation} 
which leads to a proton temperature that is
$ 10^6$~K at the coronal base, between $2\times 10^6$~K and $ 3\times 10^6 $~K in coronal holes, and $ \sim 2.5\times 10^5$~K at 1~AU. We
take the magnetic field strength to be~\citep{hollweg02}
\begin{equation}
B_0 = \left[\frac{1.5(f_{\rm max} - 1)}{d^6} + \frac{1.5}{d^2}\right] \mbox{ Gauss},
\label{eq:B0} 
\end{equation} 
with $f_{\rm max}$ (the super-radial expansion factor) equal to~5.  We
determine the rms amplitude of the fluctuating wave velocity at the
outer scale, $\delta v_{L_0} = \alpha v_{\rm A}$, using the analytical
model of \cite{chandran09c}, which describes the propagation of
low-frequency Alfv\'en waves launched outward from the Sun, taking
into account non-WKB wave reflections arising from Alfv\'en-speed
gradients as well as the cascade and dissipation of wave energy
arising from nonlinear wave-wave interactions. In particular, we set
$\delta v_{L_0}$ equal to the value of $\delta v_{\rm rms}$ plotted
with a solid line in figure~6 of \cite{chandran09c} (the curve
corresponding to their ``extended model''). We take~$L_0$ to
be~$10^4$~km at the coronal base [the limit $d\rightarrow 1$ in
equation~(\ref{eq:B0})], and to be proportional to~$B^{-1/2}$.

We consider three different values for the spectral index~$c_3$: 5/3,
3/2, and 6/5. The value $c_3 = 5/3$ is suggested by in situ
measurements of magnetic-field fluctuations in the solar
wind~\citep{matthaeus82, bruno05}, as well as some theoretical and
numerical studies of MHD turbulence~\citep{goldreich95,cho00}. The
value $c_3 =3/2$ is motivated by a different set of theoretical and
numerical studies~\citep{boldyrev06,mason06,perez09a}, as well as
recent in situ observations of the velocity power spectrum in the
solar wind~\citep{podesta07,podesta09b}. The third value, $c_3 = 1.2$,
follows from recent numerical simulations
of reflection-driven Alfv\'en-wave turbulence in coronal holes and the
fast solar wind \citep{verdini09a}.  In these simulations, $c_3= 1.2$ at~$r< 1.2
R_{\sun}$, and $c_3$ gradually increases towards~$5/3$ with
increasing~$r$.

\begin{figure*}[t]
\includegraphics[width=6.cm]{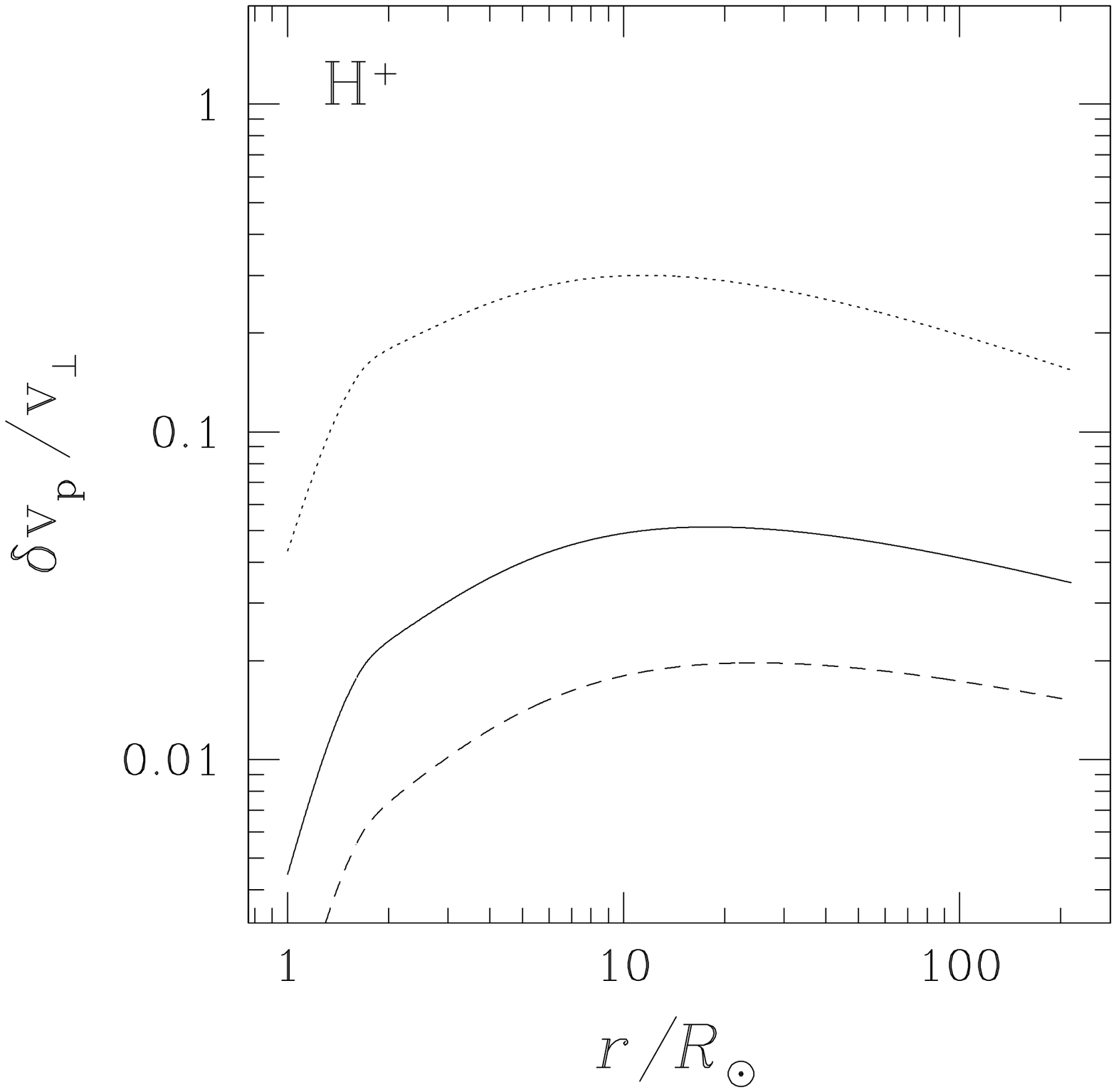}
\includegraphics[width=6.cm]{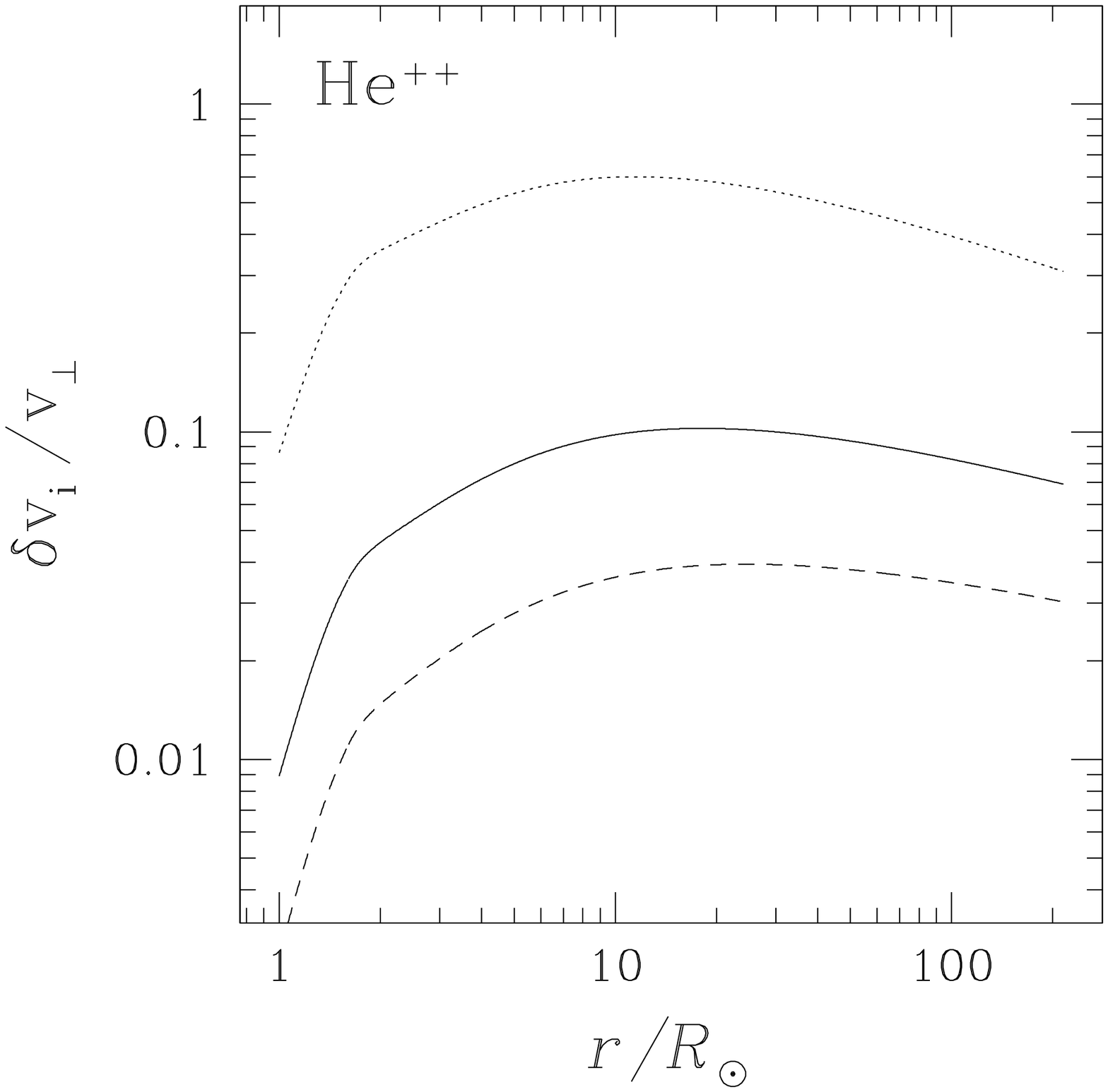}
\includegraphics[width=6.cm]{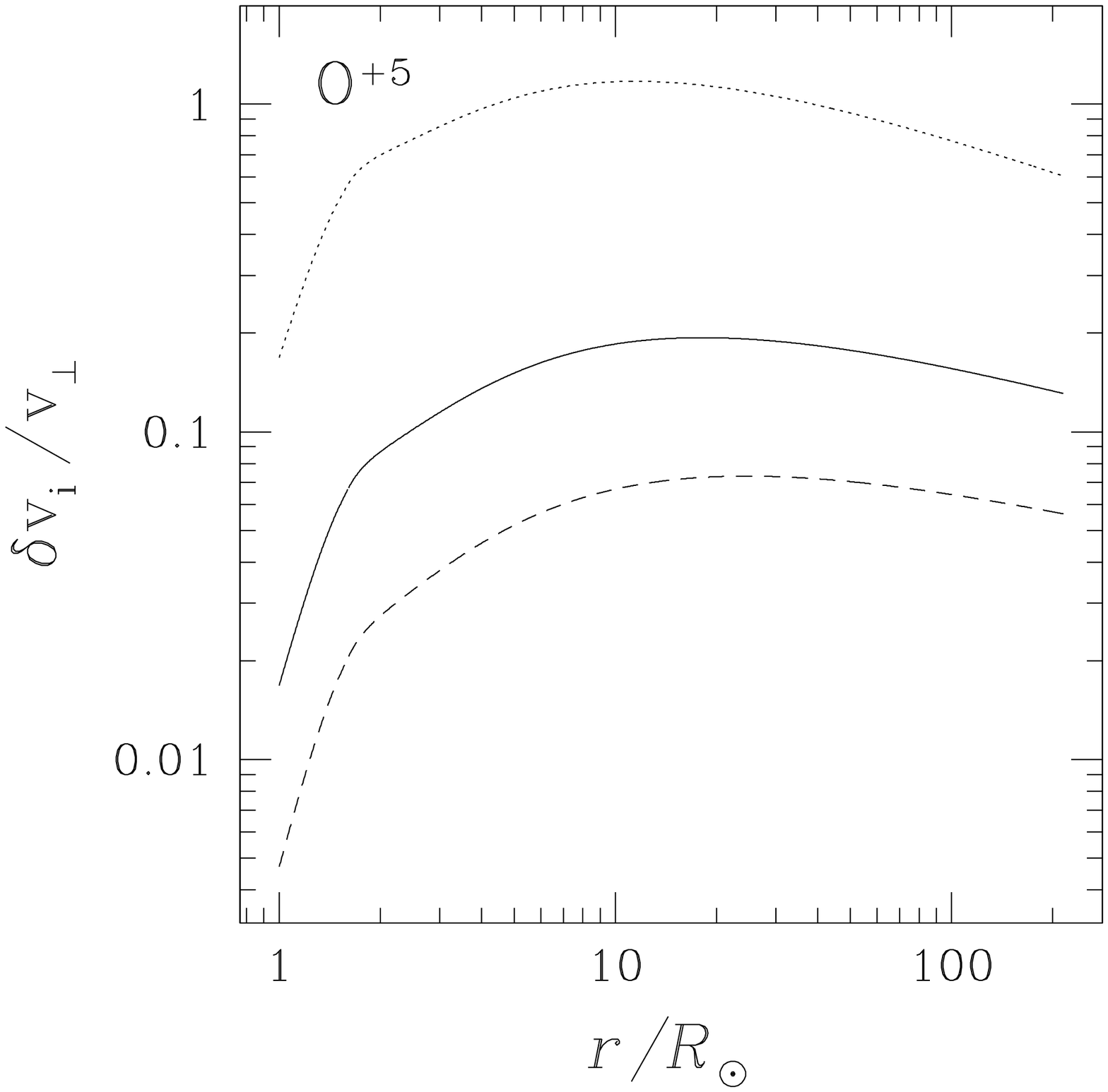}
\caption{The values of $\epsilon_{\rm i} = \delta v_{\rm i}/v_{\perp{\rm i}}$ from
  equation~(\ref{eq:eps2}) as a function of heliocentric distance for
  ${\rm H}^+$, ${\rm He}^{++}$, and ${\rm O}^{+5}$. For this figure,
  we assume that $T_\perp = T_{\rm p}$ for ${\rm He}^{++}$ and ${\rm
    O}^{+5}$ and that the one-dimensional velocity power spectrum
  $P_k^{(v)}$ is $\propto k_\perp ^{-c_3}$. From bottom to top,
  the three curves in each plot correspond to $c_3 = 5/3$, $c_3 =
  3/2$, and $c_3 = 1.2$.}\vspace{0.5cm}
\label{fig:eps_profile}
\end{figure*}

In figure~\ref{fig:eps_profile}, we plot $\epsilon_{\rm i}$ for ${\rm
  H}^+$, ${\rm He}^{++}$, and ${\rm O}^{+5}$ assuming $\mu_{\rm p} =
1.2$. Although alpha particles and minor ions are observed to be
hotter than protons in the fast solar wind, we have set all the ion
temperatures equal to~$T_{\rm p}$ to investigate the relative heating
rates of different ion species that start out at the same
temperature. Figure~\ref{fig:eps_profile} illustrates the general
point that~$\epsilon_{\rm i}$ depends strongly on the spectral index~$c_3$. In
particular, decreasing $c_3$ by 28\% from 5/3 to 1.2
increases~$\epsilon_{\rm i}$ by a factor of~$> 10$ at all radii shown for all
three ion species.  Because $Q_\perp$ depends strongly on~$\epsilon_{\rm i}$,
$Q_\perp$ is extremely sensitive to the value of~$c_3$. For example,
for protons, if $c_3 = 1.2$, then $\epsilon_{\rm p} \simeq 0.2$ except
at $r< 1.5R_{\sun}$.  The approximations leading to
equation~(\ref{eq:QG}) imply that $Q_{\perp{\rm p}}/\Gamma =
0.55$ when $\epsilon_{\rm p} = 0.2$ (where $\Gamma$ is the cascade
power at $k_\perp \rho_{\rm p} \sim 1$), indicating that perpendicular
proton heating absorbs a substantial fraction of the turbulent heating
power when~$c_3 \lesssim 1.2$.  On the other hand, if
$c_3 = 5/3$, then $\epsilon_{\rm p} < 0.02$ and $Q_{\perp {\rm
    p}}/\Gamma$ in equation~(\ref{eq:QG}) is $ < 1.2 \times
10^{-7}$.

A second general point illustrated by figure~\ref{fig:eps_profile} is
that when $c_3$ is fixed, $\epsilon_{\rm i}$ depends only weakly on~$r$
for~$2R_{\sun} < r < 1 \mbox{ AU}$.  As a result, given our
assumptions, a large radial variation in~$\epsilon$ within this range
of~$r$ requires a radial variation in the spectral index~$c_3$. As
mentioned above, the numerical simulations of \cite{verdini09a} found
$c_3 \simeq 1.2$ at $r< 1.2 R_{\sun}$, with $c_3$ increasing
towards~5/3 with increasing~$r$.  In addition, radio observations show
that the density-fluctuation power spectrum is significantly flatter
at $r= 5 R_{\sun}$ than at $r> 10
R_{\sun}$~\citep{markovskii02b,harmon05}.  These observations and
numerical simulations raise the possibility that $c_3$ is
significantly smaller (and that $Q_{\perp{\rm p}}/\Gamma$ is much larger)
close to the Sun than at $\sim 1 \mbox{ AU}$. However, the inertial
range of reflection-driven AW turbulence in coronal holes is still not
well understood. Likewise, the relation between the density power
spectrum and the velocity power spectrum in the imbalanced AW
turbulence found in coronal holes is not clear.  The $r$-dependence
of~$c_3$ thus remains uncertain.

A third point illustrated by figure~\ref{fig:eps_profile} is that
$\epsilon_{\rm i}$ is significantly larger for ${\rm He}^{++}$ and
${\rm O}^{+5}$ than for protons at the same temperature. For example,
at equal temperatures, protons and alpha particles have the same
gyroradius, and $\epsilon_\alpha = 2 \epsilon_{\rm p}$, where
$\epsilon_{\alpha}$ is the value of $\epsilon_{\rm i}$ for alpha
particles.  Because of the strong dependence of $Q_\perp$ on
$\epsilon_{\rm i}$, it is possible that the perpendicular heating rate
per unit volume from stochastic heating by gyro-scale fluctuations is
larger for alpha particles than for protons, even though Helium
comprises only~$\sim 20\%$ of the mass in the solar wind. Depending on
the values of $c_2$ and $c_3$, it is also possible that Helium absorbs
a significant fraction of the turbulent cascade power in the solar
wind. In addition, the comparatively large value of $\epsilon_{\rm i}$
for~${\rm O}^{+5}$ may explain why ${\rm O}^{+5}$ ions are observed to
be so much hotter than protons in the solar
corona~\citep{kohl98,antonucci00}, and likewise for other minor ions.

\section{Conclusion}

When an ion interacts with turbulent AWs and/or KAWs, and when the
amplitudes of the fluctuating electromagnetic fields at $\lambda_\perp
\sim \rho $ are sufficiently large, the ion's orbit becomes chaotic,
and the ion undergoes stochastic perpendicular heating.  The parameter
that has the largest effect on the heating rate is $\epsilon = \delta
v_\rho/v_\perp$, where $\delta v_\rho$ is the rms amplitude of the
velocity fluctuation at $\lambda_\perp \sim \rho $.  In the limit
$\epsilon \rightarrow 0$, the ion's magnetic moment is nearly
conserved, and perpendicular ion heating is extremely weak. On the
other hand, as $\epsilon$ increases towards unity, magnetic moment
conservation is violated, and stochastic perpendicular heating becomes
increasingly strong.

Using phenomenological arguments, we have derived an analytic formula for
the perpendicular heating rate~$Q_\perp$ for different ion species. This
formula (equation~(\ref{eq:Q2})) contains two dimensionless constants, $c_1$
and $c_2$, whose values depend on the nature of the fluctuations (e.g.,
waves versus turbulence, the slope of the power spectrum) and the shape of
the ion velocity distribution. Using test-particle simulations, we
numerically evaluate these constants for the case in which a Maxwellian
distribution of protons interacts with a spectrum of random-phase AWs and
KAWs at perpendicular wavenumbers in the range $0.264 < k_\perp \rho_{\rm p}
< 3.79$, where $ \rho_{\rm p}$ is the rms proton gyroradius in the
background magnetic field~${\bf B}_0$. The particular form of the wave power
spectrum that we choose for these simulations is motivated by the ``critical
balance'' theories of \cite{goldreich95} and \cite{cho04b}. For this case,
$c_1 = 0.75$ and $c_2 = 0.34$.  The proton heating rate~$Q_{\perp {\rm p}}$
can be compared to the cascade power~$\Gamma$ that would be present at
$k_\perp \rho_{\rm p } \sim 1$ in ``balanced'' (see
section~\ref{sec:cascade}) AW/KAW turbulence with the same value of~$\delta
v_{\rm p}$. When $c_1 = 0.75$ and $c_2 = 0.34$, the ratio~$Q_{\perp{\rm
    p}}/\Gamma$ exceeds~$1/2$ when $\epsilon _{\rm p} > \epsilon_{\rm
  crit} = 0.19$, where $\epsilon_{\rm p}$ is the value of $\epsilon$ for
thermal protons.

Our expression for $Q_{\perp {\rm p}}/\Gamma$ (equation~(\ref{eq:QG})) may
differ from the value of $Q_{\perp {\rm p}}/\Gamma$ in the solar wind for
two main reasons. First, our formula for~$\Gamma$ does not take into account
``imbalance'' (see section~\ref{sec:cascade}), which affects the relation
between~$\Gamma$ and $\delta v_{\rm p}$ in a way that is not yet
understood. Second, in true turbulence (as opposed to randomly phased
waves), a significant fraction of the cascade power may be dissipated in
coherent structures in which the fluctuating fields are larger than their
rms values~\citep{dmitruk04}.  Proton orbits in the vicinity of such
structures are more stochastic than in average regions, and thus $c_2$ may
be smaller in AW/KAW turbulence than in our test-particle simulations,
indicating stronger heating. The perpendicular heating rate is very
sensitive to the value of $c_2/\epsilon_{\rm p}$; our test-particle
simulations are consistent with $Q_{\perp {\rm p}} /\Gamma$
being~$\propto\exp(-c_2/\epsilon_{\rm p}) $. Thus, decreasing~$c_2$ leads to
a large increase in $Q_{\perp {\rm p}}/\Gamma$ when~$\epsilon_{\rm p} <
c_2$. Decreasing~$c_2$ also decreases~$\epsilon_{\rm crit}$, the value
of~$\epsilon_{\rm p}$ at which $Q_{\perp {\rm p}} /\Gamma = 1/2$; it follows
from equations~(\ref{eq:Q2}) and (\ref{eq:Gamma1}) that if $C_K \simeq 2$,
$\delta B_{\rm p} /B_0 \simeq 0.84 \delta v_{\rm p}/ v_{\rm A}$, and~$c_1
\simeq 1$, then $\epsilon_{\rm crit} \simeq c_2/2$.

When $\beta \ll 1$, stochastic proton heating by AW/KAW turbulence at
$k_\perp \rho_{\rm p} \sim 1$ increases~$T_\perp$ much more
than~$T_\parallel$. In contrast, linear proton damping of KAWs with
$\omega \ll \Omega_{\rm p}$ and $k_\perp \rho_{\rm p} \sim 1$ leads
almost entirely to parallel heating, and is only significant when the
proton thermal speed is $\gtrsim v_{\rm A}$; i.e., when $\beta_{\rm p}
\gtrsim 1$~\citep{quataert98}. If we assume that (nonlinear)
stochastic heating and linear wave damping are the only dissipation
mechanisms for low-frequency AW/KAW turbulence,\footnote{See
  \cite{markovskii02b} and \cite{markovskii06} for an argument against
  this assumption.} then we arrive at the following 
conclusions about how the cascade power in AW/KAW turbulence is partitioned between parallel and perpendicular heating, and between protons and electrons:
\begin{enumerate}
\item If $\beta_{\rm p} \ll 1 $ and $\epsilon_{\rm p}\ll \epsilon_{\rm
    crit}$, then proton heating is negligible and electrons
  absorb most of the cascade power.
\item If $\beta_{\rm p}\ll 1$ and $\epsilon_{\rm p} \gtrsim
  \epsilon_{\rm crit}$, then parallel proton heating is negligible,
  and AW/KAW turbulence leads to a combination of electron heating and
  perpendicular proton heating.
\item If $\beta_{\rm p} \gtrsim 1$ and $\epsilon_{\rm p} \ll
  \epsilon_{\rm crit}$, then perpendicular proton heating is
  negligible, and AW/KAW turbulence results in a combination of
  electron heating and parallel proton heating.
\item If $\beta_{\rm p} \gtrsim 1$ and $\epsilon_{\rm p} \gtrsim
  \epsilon_{\rm crit}$, then perpendicular proton heating, parallel
  proton heating, and electron heating each receives an appreciable
  fraction of the cascade power.
\end{enumerate} 
KAW turbulence at $k_\perp \rho_{\rm p} \sim 1$ fluctuates over length
(time) scales much greater than $\rho_{\rm e}$ ($\Omega_{\rm
  e}^{-1}$), where $\rho_{\rm e}$ ($\Omega_{\rm e}$) is the
thermal-electron gyroradius (cyclotron frequency). Because of this, an
electron's magnetic moment is nearly conserved when it interacts with
KAW turbulence at $k_\perp \rho_{\rm p} \sim 1$. Electron heating by
KAW turbulence at $k_\perp \rho_{\rm p} \sim 1$ is thus primarily
parallel heating. On the other hand, some of the fluctuation energy
may cascade to scales~$\ll \rho_{\rm p}$. The way that turbulence is
dissipated at such scales is not yet well understood.

To determine the dependence of~$\epsilon_{\rm i}$ (the value
of~$\epsilon$ for thermal ions) on heliocentric distance~$r$ for
different ion species in the fast solar wind, we adopt a simple
analytic model for the radial profiles of the solar-wind proton
density, proton temperature, and magnetic field strength. We then
apply the analytical model of \cite{chandran09c}, which describes the
radial dependence of the rms amplitudes of Alfv\'en waves at the outer
scale~$L_0$ of the turbulence, and assume that the velocity power
spectrum~$P_k^{(v)}$ is~$\propto k_\perp^{-c_3}$ for $L_0^{-1} <
k_\perp < \rho_{\rm p}^{-1}$.  We find that the value of
$\epsilon_{\rm i}$ for protons, Helium, and minor ions depends
strongly on~$c_3$. However, for a fixed value of~$c_3$, $\epsilon_{\rm
  i}$ is relatively insensitive to~$r$ for $2R_{\sun} < r < \mbox{ 1
  AU}$.

We are not yet able to determine with precision the perpendicular heating
rates of different ion species as a function of~$r$ because of the
uncertainties in the values of~$c_2$ and~$c_3$ in the solar wind, and because of the large sensitivity of the heating rates to these quantities.
However, if we assume that the value of $c_2$ for protons in the solar
wind is close to the value of~0.34 in our test-particle simulations,
then we arrive at the following two conclusions. First, perpendicular
proton heating is a negligible fraction of the turbulent cascade power
in the bulk of the explored solar wind, in which $c_3$ is measured to
be in the range of 1.5 - 1.7. Second, if stochastic proton heating is
important close to the Sun, then $c_3$ must be significantly smaller
close to the Sun than at 1~AU. (For example, if $c_3 = 1.2$, then
$Q_{\perp{\rm p}}/\Gamma \gtrsim 0.5$ for $ 2 R_{\sun}\lesssim r < 100
R_{\sun}$.)  

We find that alpha particles and minor ions undergo much stronger
stochastic heating than protons, in large part part because the value
of~$\epsilon_{\rm i}$ is larger for these ions than for protons at
equal temperatures. Depending on the values of $c_2$ and~$c_3$, the
stochastic heating rate per unit volume in the solar wind may be
larger for Helium than for protons, even though Helium comprises
only~$\sim 20\%$ of the solar-wind mass.  Figure~\ref{fig:eps_profile}
suggests that stochastic heating is important for alpha particles and
minor ions even if $c_3$ is as large as~3/2, since $\epsilon_{\rm i}$
is then $\gtrsim 0.1$ over a wide range of~$r$.  However, further
investigations into the value of~$c_3$ close to the Sun and the value
of~$c_2$ for (non-random-phase) AW/KAW turbulence are needed in order
to develop a more complete and accurate picture of stochastic ion
heating in the solar wind.

\acknowledgements We thank Greg Howes for providing us with the
numerical value of the constant~$C_{\rm K}$ that appears in
equation~(\ref{eq:Gamma1}).  This work was supported in part by the
Center for Integrated Computation and Analysis of Reconnection and
Turbulence (CICART) under DOE Grant DE-FG02-07-ER46372, and by NSF
Grant ATM-0851005, NSF-DOE Grant AST-0613622, and NASA Grants
NNX07AP65G and NNX08AH52G.  E.~Q.\ was supported in part by NSF-DOE
Grant PHY-0812811, NSF Grant ATM-0752503, the David and Lucille
Packard Foundation, and the Miller Institute for Basic Research in
Science, University of California Berkeley.

\appendix

\section{Leading-Order Conservation of the First Adiabatic Invariant When $k_\perp \rho  \sim 1$ and $\epsilon \ll 1$}
\label{ap:mu} 

\begin{figure}[h]
   \centerline{\includegraphics[width=6.cm]{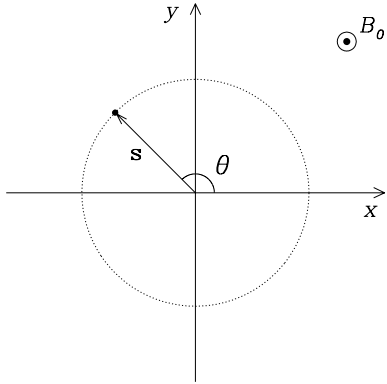}}
\caption{In the small-$\epsilon$ limit, an ion's trajectory in the $xy$ plane is approximately a circle centered on its guiding-center position~${\bf R}$.}
\label{fig:gyro_average}
\end{figure}

In this appendix, we consider the interaction between ions and low-frequency, 2D
($k_\parallel = 0$), electrostatic fluctuations with $k_\perp \rho \sim 1$. We assume that $\epsilon \ll 1$, neglect magnetic-field
fluctuations, and show that the leading-order non-vanishing terms in
$dH/dt$ are unable to cause secular perpendicular ion heating.
We set ${\bf B} = B_0 {\bf \hat{z}}$, where $B_0$ is a constant. The
time derivative of the ion's guiding-center position, defined in
equation~(\ref{eq:defR}), is then given by
\begin{equation}
\frac{d{\bf R}}{dt} = v_z {\bf \hat{z}} + \frac{c {\bf E} \times \hat{z}}{B_0}.
\label{eq:dRdt2} 
\end{equation} 
Since $\epsilon \ll 1$, the particle's orbit in the
$xy$-plane during a single gyroperiod is approximately a circle of
radius~$\rho  = v_\perp/\Omega  $.  
We assume that $\Phi$ varies slowly
in time, on a time scale of~$\sim \epsilon^{-1} \Omega  ^{-1}$, with $\partial \Phi/\partial z = 0$.
We introduce two related forms of ``gyro-averages.'' First,
if $h$ is some physical property of a particle, such as its energy or guiding-center velocity, then
we define the gyro-average of~$h$ to be 
\begin{equation}
\langle h(t) \rangle  = \frac{\Omega  }{2\pi} \int_{t- \pi/\Omega  }
^{t+\pi/\Omega  } h(t_1) \,dt_1.
\label{eq:gyro2} 
\end{equation}
Second, for a general function of position and time $g({\bf r}, t)$ satisfying $\partial g/\partial z = 0$, we define
the gyro-average of $g({\bf r}, t)$ 
for particles with perpendicular velocity~$v_\perp$ and guiding center~${\bf R}$ 
to be given by
\begin{equation}
\langle g({\bf r},t)\rangle_{{\bf R}, v_\perp} \equiv 
\frac{1}{2\pi} \int_0^{2\pi}  g  \mbox{\large \bf (}{\bf R} + {\bf s}(\theta), t 
\mbox{\large \bf )}\, d\theta,
\label{eq:gyro} 
\end{equation} 
where ${\bf s} =  {\bf \hat{x}} \rho \cos(\theta) + {\bf \hat{y}}\rho  \sin(\theta)$ is the vector illustrated in
figure~\ref{fig:gyro_average}.

To simplify the notation, we define
\begin{equation}
\overline{ g}({\bf R}, t) \equiv \langle g({\bf r},t)\rangle_{{\bf R}, v_\perp},
\label{eq:defgbar} 
\end{equation} 
where the functional dependence of~$\overline{ g}$ on $v_\perp$ is
not explicitly written.  If $g$ varies slowly in time at a
fixed point in space (e.g., on the time scale~$\epsilon^{-1}
\Omega  ^{-1}$), then $\overline{ g}({\bf R}, t)$ is (to leading
order in~$\epsilon$) equivalent to a time average over one cyclotron
period of~$g({\bf r}, t)$ evaluated at the position~${\bf r}(t)$ of a
particle with guiding center~${\bf R}$:
\begin{equation}
\overline{ g}({\bf R},t)  = \frac{\Omega}{2\pi}\int_{t-\pi/\Omega}^{t+\pi/\Omega} g\mbox{\large \bf (}{\bf r}(t_1), t_1\mbox{\large \bf )} \,dt_1.
\label{eq:gyro0} 
\end{equation} 
Thus, if we take the gyro-average of the ``particle property'' $d{\bf R}/dt$ 
in equation~(\ref{eq:dRdt2})  using equation~(\ref{eq:gyro2}), we find that
\begin{equation}
 \left\langle \frac{d{\bf R}}{dt} \right\rangle
= v_{\rm z} {\bf \hat{z}} + \frac{c}{B_0} \langle {\bf E} ({\bf r}, t)\rangle_{{\bf R}, v_\perp}\times {\bf \hat{z}}  .
\label{eq:dRdt3} 
\end{equation} 
We consider electrostatic fluctuations with $\partial {\bf A}/\partial t
= 0$, and thus, ${\bf E} = -\nabla \Phi$.
Omitting the explicit time dependence of~$\Phi$ and $\overline{ \Phi}$ to simplify the notation,  we can write
the gyro-average of $\partial \Phi/\partial x$ as
\begin{equation}
\left\langle \frac{\partial \Phi}{\partial x} \right\rangle_{{\bf R},v_\perp} = \lim_{\delta \rightarrow 0}
\left\langle \frac{\Phi({\bf r} + {\bf \hat{x}} \delta) - \Phi({\bf r})}{\delta }\right\rangle_{{\bf R},v_\perp}
= \lim_{\delta \rightarrow 0 } \frac{\overline{\Phi}({\bf R} + {\bf \hat{x}}\delta ) -
\overline{ \Phi}({\bf R})}{\delta } = \frac{\partial \overline{ \Phi}}{\partial x^\prime}
\label{eq:gyro_dphidx} 
\end{equation} 
where ${\bf \hat{x}}$ is a unit vector in the~$x$ direction, and
$\partial/\partial x^\prime$ denotes a partial derivative with respect
to the x-component of the guiding-center position~${\bf R}$.  
Equation~(\ref{eq:dRdt3}) can thus be re-written as
\begin{equation}
 \left\langle \frac{d{\bf R}}{dt} \right\rangle
= v_{\rm z} {\bf \hat{z}} - \frac{c}{B_0} \nabla^\prime \overline{ \Phi}\times {\bf \hat{z}}  ,
\label{eq:vD} 
\end{equation} 
where $\nabla^\prime$ indicates a gradient with respect to the coordinates
of the guiding-center position~${\bf R}$.  Since we have assumed
$\partial /\partial z = 0$, equation~(\ref{eq:vD}) implies that $
\langle d{\bf R}/dt \rangle \cdot \nabla^\prime \overline{ \Phi} = 0$.

We now integrate equation~(\ref{eq:dHdt}) for an integral number of
cyclotron periods, from $t_{\rm a}$ to $t_{\rm b} = t_{\rm a} + N
\delta t$, where $\delta t = 2\pi/\Omega  $ and $N\geq
\epsilon^{-1}$. We define $t_0 = t_{\rm a} + \delta t/2$  and $t_j = t_{j-1} + \delta t$ for any integer~$j$.  Since
we have assumed that $\partial {\bf A}/\partial t = 0$, the integral
of equation~(\ref{eq:dHdt}) can be written
\begin{equation}
H(t_{\rm b}) - H(t_{\rm a}) =  q\sum_{j=0}^{N-1}\int_{t_j-\delta t/2}^{t_j+\delta t/2} \frac{\partial \Phi}{\partial t}\,dt  .
\label{eq:integ1} 
\end{equation} 
In analogy to equation~(\ref{eq:gyro_dphidx}), it is straightforward to show that
$\overline{  \partial \Phi/\partial t} = (\partial/\partial t) \overline{\Phi}$. 
We can thus re-write equation~(\ref{eq:integ1}) as
\begin{equation}
H(t_{\rm b}) - H(t_{\rm a}) =  q\sum_{j=0}^{N-1}  \,\frac{\partial \overline{\Phi} }{\partial t} \mbox{\large \bf (}{\bf R}(t_j),t_j\mbox{\large \bf )}\, \delta t.
\label{eq:integ2} 
\end{equation} 
The time scale on which
$\overline{\Phi} \mbox{\large \bf (}{\bf R}(t),t\mbox{\large \bf )}$ changes by a factor of order unity
is~$\epsilon^{-1} \delta t$. This is because $\Phi$ changes slowly in time at a fixed point in space,
$k_\perp \rho  \sim 1$,  and
$d{\bf R}/dt \sim \epsilon v_\perp$.  As a result $\partial \overline{
  \Phi}/\partial t$ is approximately constant within each time
interval of duration~$\delta t$. The right-hand side of
equation~(\ref{eq:integ2}) is therefore a discrete approximation of the
integral of $q\partial \overline{ \Phi}/\partial t$ from $t_{\rm a}$
to $t_{\rm b}$, with a fractional error of order~$\epsilon$, so that
\begin{equation}
H(t_{\rm b}) - H(t_{\rm a}) =  q\int_{t_{\rm a}}^{t_{\rm b}} 
\frac{\partial }{\partial t}\, \overline{\Phi}  \mbox{\large \bf (}{\bf R}(t),t\mbox{\large \bf )}\, dt
+ \dots,
\label{eq:integ3} 
\end{equation} 
where the ellipsis ($ \dots$) represents corrections
that are higher order in~$\epsilon$.
The right-hand side of equation~(\ref{eq:integ3}) can be re-written in terms of the total time derivative of~$\Phi$, yielding
\begin{equation}
H(t_{\rm b}) - H(t_{\rm a}) =  q\int_{t_{\rm a}}^{t_{\rm b}} 
\frac{d}{d t}\, \overline{\Phi}  \mbox{\large \bf (}{\bf R}(t),t\mbox{\large \bf )}\, dt
\: -\:
q\int_{t_{\rm a}}^{t_{\rm b}}  \frac{d{\bf R}}{dt} \cdot \nabla^\prime
\overline{\Phi}  \mbox{\large \bf (}{\bf R}(t),t\mbox{\large \bf )}\, dt
+ \dots
\label{eq:integ3.25} 
\end{equation} 
Since $\nabla^\prime \overline{  \Phi}$  is nearly constant during a single time interval of duration~$\delta t$, the second integral on the right-hand side of equation~(\ref{eq:integ3.25}) 
satisfies the relation
\begin{equation}
\int_{t_{\rm a}}^{t_{\rm b}}  \frac{d{\bf R}}{dt} \cdot \nabla^\prime
\overline{\Phi}  \mbox{\large \bf (}{\bf R}(t),t\mbox{\large \bf )}\, dt
= \sum_{j=0}^{N-1}\left(\int_{t_j-\delta t/2}^{t_j+\delta t/2}  \frac{d{\bf R}}{dt}\,dt  \right)\cdot \nabla^\prime \overline{\Phi}  \mbox{\large \bf (}{\bf R}(t_j),t_j\mbox{\large \bf )} + \dots
\label{eq:integ3.5} 
\end{equation} 
The integral within parentheses 
on the right-hand side of equation~(\ref{eq:integ3.5}) is equivalent to $
\langle d{\bf R}/dt\rangle \, \delta t$ evaluated at $t=t_j$.
From equation~(\ref{eq:vD}),   
 $\langle d{\bf R}/dt\rangle\cdot
\nabla^\prime \overline{ \Phi} = 0 $. Thus, the right-hand side of equation~(\ref{eq:integ3.5}) and the second integral on the right-hand side of equation~(\ref{eq:integ3.25})   vanish to leading order in~$\epsilon$. Equation~(\ref{eq:integ3.25}) thus becomes
\begin{equation}
H(t_{\rm b}) - H(t_{\rm a}) =  q \overline{ \Phi}\mbox{\large \bf (}{\bf R}(t_{\rm b}),t_{\rm b}\mbox{\large \bf )}
- q \overline{ \Phi}\mbox{\large \bf (}{\bf R}(t_{\rm a}),t_{\rm a}\mbox{\large \bf )} + \dots
\label{eq:integ6} 
\end{equation} 
The right-hand side of equation~(\ref{eq:integ6}) remains $\lesssim
q\delta \Phi_\rho$, regardless of how large the interval $(t_{\rm b} - t_{\rm a})$ becomes. 
Thus, to leading order in~$\epsilon$, there is no secular change in the particle
energy~$H$, consistent with the near-conservation of the first
adiabatic invariant in the small-$\epsilon$, small-$\omega/\Omega  $
limits.

\bibliography{articles}
\bibliographystyle{jphysicsB}

\end{document}